\documentclass[11pt]{article}
\usepackage{epsfig} 
\newcommand{\sect}[1]{ \section{#1} \setcounter{equation}{0} }

\newcommand{\pslash}{p \! \! \! /} 
\newcommand{\qslash}{q \! \! \! /}

\newcommand{\half}{\mbox{\small{$\frac{1}{2}$}}}

\newcommand{\MSbar}{\overline{\mbox{MS}}} 
\newcommand{\MSbars}{\overline{\mbox{\footnotesize{MS}}}} 
\newcommand{\Nf}{N_{\!f}}

\newcommand{\NA}{N_{\!A}}

\begin{document}
\title{RI${}^\prime$/SMOM scheme amplitudes for deep inelastic scattering 
operators at one loop in QCD}
\author{J.A. Gracey, \\ Theoretical Physics Division, \\ 
Department of Mathematical Sciences, \\ University of Liverpool, \\ P.O. Box 
147, \\ Liverpool, \\ L69 3BX, \\ United Kingdom.} 
\date{} 
\maketitle 

\vspace{5cm} 
\noindent 
{\bf Abstract.} We compute the amplitudes for the insertion of various
operators in a quark $2$-point function at one loop in the RI${}^\prime$
symmetric momentum scheme, RI${}^\prime$/SMOM. Specifically we focus on the 
moments $n$~$=$~$2$ and $3$ of the flavour non-singlet twist-$2$ operators used
in deep inelastic scattering as these are required for lattice computations.

\vspace{-15.5cm}
\hspace{13.5cm}
{\bf LTH 885}

\newpage

\sect{Introduction.}

Lattice gauge theories are the main theoretical tool for exploring the 
non-perturbative r\'{e}gime of the strong nuclear force by simulating the
underlying quantum field theory which is the non-abelian gauge theory of quarks
and gluons, (QCD). The numerical techniques allows one to explore the infrared 
region where perturbation theory becomes impractical because the value of the 
parameter controlling the expansion, which is the coupling constant, becomes 
large. Briefly, one uses the path integral formalism but with the space-time 
continuum replaced by a discrete Euclidean grid. Then one can construct Green's
functions numerically, representing, for example, bound state particle spectra,
and make measurements of the masses. Whilst overlooking many of the technical
aspects of this procedure at this point, it is a remarkable achievement that 
the formalism is in more than solid agreement with nature. Aside from 
determining particle spectra, one of the current main problems is to measure 
matrix elements of operators in the non-perturbative region. These are 
important in, for instance, understanding the structure of nucleons if one 
focuses on the operators relating to deep inelastic scattering. These were 
introduced originally in \cite{1} which subsequently produced an intense 
industry to determine the operator anomalous dimensions for arbitrary moment to
eventually three loops in the $\MSbar$ renormalization scheme, 
\cite{2,3,4,5,6,7}. Indeed there has been a large degree of progress in the 
area of measuring matrix elements of quark bilinear currents and operators and 
the associated renormalization constants by the QCDSF collaboration, 
\cite{8,9,10,11,12,13,14}, and others \cite{15,16,17,18,19,20,21,22,23,24}.
However, in order to make reliable measurements and hence accurate predictions 
one has to overcome various theoretical difficulties. Aside from those relating 
specifically to the lattice, there is the problem of ensuring the results match
on to what would be expected at high energies. In other words the Green's 
function depends on some reference momentum value and the numerical 
simulations, in principle, make measurements not only at low but also at high 
energies. In the latter case perturbation theory is actually valid there and 
hence is a reliable complementary tool. Therefore, if the same Green's function
is computed to several loop orders then it ought to be the case that the 
numerical measurements will overlap at large energies. Given this, it is 
sometimes the situation when there are accurate large loop order results that 
the continuum estimate is used to assist with normalizing the lattice 
measurements.

This brief overview clouds some of the more technical aspects of the overall
procedure. For instance, all the operators of interest undergo renormalization.
Whilst this is not a major problem, since the formalism to carry out a 
renormalization has been established for many years now, there is the problem
of the relation of a continuum renormalization to what is performed in practice
on the lattice. For instance, the standard practice in high energy problems is 
to dimensionally regularize QCD and then to subtract the resulting divergences,
which are manifested as poles in the regularizing parameter $\epsilon$, in a 
minimal or modified minimal way. The latter scheme, $\MSbar$, is the main 
procedure primarily as convergence is improved by removing a specific finite 
part in addition to the basic poles. The main advantage of this mass 
independent renormalization scheme is that one can compute to very high loop 
order when the quarks are massless and even to a reasonable order in some cases
when there are massive quarks. Whilst this provides accurate results for the 
lattice to match to, there is a technical problem to be overcome which is that
the lattice computations are invariably in a non-minimal renormalization 
scheme. So to make a proper comparison for matching at high energy one has to 
convert the results to the {\em same} scheme. One of the more widely used 
lattice schemes is the RI${}^\prime$ scheme, \cite{25,26}, which denotes the 
modified regularization invariant scheme. It is a modification of the RI 
scheme, \cite{25,26}, where the essential difference is in the way the quark
wave function renormalization constant is defined. Briefly, the difference 
between the RI${}^\prime$ and RI schemes is in a differentiation of the quark
$2$-point function with respect to the momentum. As the derivative has a 
financial cost for the lattice, the RI${}^\prime$ scheme is more efficient and 
hence is the default scheme in this respect, \cite{25,26}. Both these lattice 
schemes are mass dependent renormalization schemes and are a hybrid of $\MSbar$
and MOM schemes. By this we mean that in the main $2$-point functions are 
renormalized according to a MOM type subtraction whilst three and higher point 
functions are renormalized using $\MSbar$. This ensures, for example, that the 
RI and RI${}^\prime$ scheme coupling constants are the same as that of the 
$\MSbar$ scheme, \cite{25,26}. Whilst introduced in \cite{25,26}, the 
renormalization of QCD in the continuum has been studied in the Landau gauge in
\cite{27} and later in a general linear covariant gauge in \cite{28}. 

Consequently, with interest in measuring matrix elements on the lattice 
relating to nucleon structure there has been a need to carry out the continuum
RI${}^\prime$ renormalization of the same Green's functions. These involve the
low moment flavour non-singlet twist-$2$ Wilson operators which arise in the
deep inelastic scattering formalism. Indeed three loop results are available in
RI${}^\prime$ in \cite{28,29,30}. As these renormalization constants are 
determined by inserting the operator into a quark $2$-point function with 
massless quarks then one could apply standard algorithms such as {\sc Mincer},
\cite{31,32}, to achieve high loop orders. However, from a technical point of 
view this renormalization is carried out at a point of exceptional momentum 
since the operator is inserted at zero momentum. In this configuration one is 
effectively dealing with a reduced $2$-point function. It has recently been 
pointed out, \cite{33}, that for matrix elements relating to quark masses then 
one could have infrared issues in extracting reliable results in the 
non-perturbative region as a consequence of this momentum configuration. To 
circumvent this technicality an alternative scheme has been developed which is 
called the RI${}^\prime$/SMOM scheme, \cite{33}. It differs from the 
RI${}^\prime$ scheme in that $3$-point functions are not subtracted at an 
exceptional point but instead at a symmetric point. This means that none of the
external momenta are nullified, so that the potential infrared singularity 
embedded within the logarithms of the Green's function are bypassed, \cite{33}.
Whilst the one loop computations were performed in \cite{33}, this has recently
been extended to two loops in \cite{34,35} for the scalar and tensor currents.
However, given that there is recent interest in measuring twist-$2$ flavour
non-singlet operator matrix elements on the lattice for low moments, \cite{23},
it is the purpose of this article to present the first one loop computations 
for the moments $n$~$=$~$2$ and $3$. Although lattice computations focus on the
Landau gauge, partly as that gauge is less complicated to fix numerically, the 
Green's function with the operator insertion is gauge dependent. So in 
developing another renormalization scheme to overcome one technical problem 
there are potential numerical errors in measurements from gauge fixing issues. 
Instead all our results will be in a general linear covariant gauge. Whilst the
renormalization of these Wilson operators has already been carried out for the 
RI${}^\prime$ scheme, \cite{28,29,30}, it may appear to be rather simple to 
follow that procedure for the latest scheme. This is not the case for an 
elementary reason. This is to do with the fact that the Wilson operators mix 
under renormalization, \cite{36}. It is widely accepted that the flavour
singlet Wilson operators mix among themselves and the flavour non-singlet ones 
do not, \cite{1,2,3}. Indeed in the original context of \cite{1,2,3} this is
the situation. However, that is only the case for the latter set if the Green's
function containing the operator does not have a momentum flowing out of the 
operator itself. If there is a net momenta flow through the operator then they 
mix into a set of total derivative operators. Given that the RI${}^\prime$/SMOM
scheme is at non-exceptional momenta then this mixing cannot be ignored. It has
been studied to three loops in a practical situation in \cite{36} where a 
similar problem for the lattice was examined but in a context which involves a 
Green's function which is gauge independent but contains two operators. Indeed
the mixing matrix for the two Wilson operators we consider here was calculated 
to three loops in the $\MSbar$ scheme. Without including the mixing in 
\cite{36} the operator correlation function did not correctly satisfy the 
renormalization group equation at two and three loops.

The article is organised as follows. We review the problem and the necessary
background in the following section. All our results are collected in sections
three and four where we give all the amplitudes as a function of the gauge 
parameter of an arbitrary linear covariant gauge for both the $\MSbar$ and 
RI${}^\prime$/SMOM schemes. We conclude in section five and include an
appendix. It contains the bases of tensors into which all the Green's functions
are decomposed together with the coefficients of the projection tensors which 
project out each specific amplitude. 

\sect{Preliminaries.}

In this section we recall the background to the problem and the calculational
set-up. First, the various operators we will be considering, which are gauge
invariant, are
\begin{eqnarray}
S & \equiv & \bar{\psi} \psi ~, \nonumber \\
V & \equiv & \bar{\psi} \gamma^\mu \psi ~, \nonumber \\
T & \equiv & \bar{\psi} \sigma^{\mu\nu} \psi ~, \nonumber \\
W_2 & \equiv & {\cal S} \bar{\psi} \gamma^\mu D^\nu \psi ~, \nonumber \\
\partial W_2 & \equiv & {\cal S} \partial^\mu \left( \bar{\psi} \gamma^{\nu}
\psi \right) ~, \nonumber \\
W_3 & \equiv & {\cal S} \bar{\psi} \gamma^\mu D^\nu D^\sigma \psi 
\nonumber \\
\partial W_3 & \equiv & {\cal S} \partial^\mu \left( \bar{\psi} \gamma^\nu
D^\sigma \psi \right) ~, \nonumber \\
\partial \partial W_3 & \equiv & {\cal S} \partial^\mu \partial^\nu \left(
\bar{\psi} \gamma^\sigma \psi \right) 
\label{opdef}
\end{eqnarray}
where the first three operators are included for checking purposes and all 
derivatives, both ordinary and covariant, act to the right. In (\ref{opdef}) 
${\cal S}$ means total symmetrization in the free Lorentz indices and we use 
the same labelling and notation as \cite{36} for ease of reference. For 
instance, at certain points in this respect we will refer to the level $W_2$ or
$W_3$. By this we will mean either the specific operator with that label or the
set of operators within that level which are additionally either $\partial W_2$
or $\partial W_3$ and $\partial \partial W_3$ respectively. Like \cite{36}, it 
will be clear from the context which is meant. As discussed already one must 
include these additional total derivative operators within each level since 
there is mixing between the operators. Such mixings must be included when there
is a momentum flowing through the operator insertion in the Green's function 
irrespective of the number of such included operators. For the earlier 
RI${}^\prime$ scheme computations of the anomalous dimensions the mixing issue 
was not relevant since the operator was inserted at zero momentum, 
\cite{28,29,30}. Given that we are considering operators with free Lorentz 
indices, whether they relate to deep inelastic scattering or not, we cannot 
follow the earlier prescription of \cite{1,2,3}. There the free Lorentz indices
of the matrix elements were contracted with a null vector, $\Delta_\mu$, which 
projected out that part of the matrix element containing the divergence. The 
reason that we have to take a different approach resides in the fact that on 
the lattice measurements are made for various individual components of the free
indices. Therefore, we have to take a more general approach and decompose our
Green's functions into a basis of Lorentz tensors which have the {\em same}
symmetry structure as the operator which is inserted into the Green's function.
For the tensor current this means that the basis has to be antisymmetric in
the two free indices since 
$\sigma^{\mu\nu}$~$=$~$\half[\gamma^\mu ,\gamma^\nu]$. In the case of the two 
Wilson operators each operator in the respective levels are totally symmetric 
in the indices and are traceless in $d$-dimensions, \cite{1}. To be specific we
have 
\begin{eqnarray}
{\cal S} {\cal O}^{W_2}_{\mu\nu} &=& {\cal O}^{W_2}_{\mu\nu} ~+~
{\cal O}^{W_2}_{\nu\mu} ~-~ \frac{2}{d} \eta_{\mu\nu}
{\cal O}^{W_2\,\sigma}_{\sigma} ~, \nonumber \\
{\cal S} {\cal O}^{W_3}_{\mu\nu\sigma} &=&
{\cal O}^{W_3}_{S~\mu\nu\sigma} ~-~ \frac{1}{(d+2)} \left[
\eta_{\mu\nu} {\cal O}^{W_3~~\rho}_{S~\sigma\rho} ~+~
\eta_{\nu\sigma} {\cal O}^{W_3~~\rho}_{S~\mu\rho} ~+~
\eta_{\sigma\mu} {\cal O}^{W_3~~\rho}_{S~\nu\rho} \right] 
\end{eqnarray}
with 
\begin{eqnarray}
{\cal O}^{W_3}_{S~\mu\nu\sigma} &=&
\frac{1}{6} \left[ {\cal O}^{W_3}_{\mu\nu\sigma} ~+~
{\cal O}^{W_3}_{\nu\sigma\mu} ~+~
{\cal O}^{W_3}_{\sigma\mu\nu} ~+~
{\cal O}^{W_3}_{\mu\sigma\nu} ~+~
{\cal O}^{W_3}_{\sigma\nu\mu} ~+~
{\cal O}^{W_3}_{\nu\mu\sigma} \right] 
\end{eqnarray}
where the basic operators are 
\begin{eqnarray}
{\cal O}^{W_2}_{\mu\nu} &=& \bar{\psi} \gamma_\mu D_\nu \psi ~, \nonumber \\
{\cal O}^{W_3}_{\mu\nu\sigma} &=& \bar{\psi} \gamma_\mu D_\nu D_\sigma \psi 
\end{eqnarray}
and $D_\mu$ is the usual covariant derivative. These are the same definitions
as used in earlier computations, \cite{28,29,30}. The total derivative 
operators in the same respective levels satisfy these same template 
definitions. (We have suppressed the free flavour indices but note that these 
are flavour non-singlet operators.) 

Next we recall the key points concerning the mixing of the operators in levels
$W_2$ and $W_3$. First, as we are working with massless quarks there are no
lower dimensional operators to be included and there is no mixing between
levels. Next the particular choice of operators, (\ref{opdef}), means that 
whilst there is mixing the mixing matrix of renormalization constants is upper
triangular and given by, \cite{36}, 
\begin{equation}
Z^{W_2}_{ij} ~=~ \left(
\begin{array}{cc}
Z^{W_2}_{11} & Z^{W_2}_{12} \\
0 & Z^{W_2}_{22} \\
\end{array}
\right)
\label{mat2}
\end{equation}
and 
\begin{equation}
Z^{W_3}_{ij} ~=~ \left(
\begin{array}{ccc}
Z^{W_3}_{11} & Z^{W_3}_{12} & Z^{W_3}_{13} \\
0 & Z^{W_3}_{22} & Z^{W_3}_{23} \\
0 & 0 & Z^{W_3}_{33} \\
\end{array}
\right) ~.
\label{mat3}
\end{equation}
Again we avoid a clumsy index on the matrix elements by using a numerical map
to the respective sets $\{W_2,\partial W_2\}$ and $\{W_3,\partial W_3,\partial 
\partial W_3\}$ respectively. Here the superscript indicates the level. Once 
these have been determined in a specific renormalization scheme then the 
anomalous dimension matrix is deduced from  
\begin{equation}
\gamma^{\cal O}_{ij} ~=~ \mu \frac{d ~}{d \mu}
\ln Z^{\cal O}_{ij}
\label{mixmatdef}
\end{equation}
with
\begin{equation}
\mu \frac{d~}{d\mu} ~=~ \beta(a) \frac{\partial ~}{\partial a} ~+~
\alpha \gamma_\alpha(a,\alpha) \frac{\partial ~}{\partial \alpha} 
\label{muderiv}
\end{equation}
where $\alpha$ is the gauge parameter of the canonical linear covariant gauge
and $\alpha$~$=$~$0$ corresponds to the Landau gauge. Although the operators
we are considering are gauge invariant, in a mass dependent renormalization 
scheme, such as RI${}^\prime$ or RI${}^\prime$/SMOM, the anomalous dimensions 
can depend on the gauge. This is why we have included the second term on the 
right side of (\ref{muderiv}). However, for the one loop computation here the 
leading term is scheme independent so that there is no gauge dependence at this
order. In order to compare with the structure of our RI${}^\prime$/SMOM results
later, we recall the three loop $\MSbar$ scheme anomalous dimension mixing 
matrices, \cite{36}, are 
\begin{eqnarray}
\gamma^{W_2}_{11}(a) &=& \frac{8}{3} C_F a ~+~ \frac{1}{27} \left[ 376 C_A C_F
- 112 C_F^2 - 128 C_F T_F \Nf \right] a^2 \nonumber \\
&& +~ \frac{1}{243} \left[ \left( 5184 \zeta(3) + 20920 \right) C_A^2 C_F
- \left( 15552 \zeta(3) + 8528 \right) C_A C_F^2 \right. \nonumber \\
&& \left. ~~~~~~~~~~-~ \left( 10368 \zeta(3) + 6256 \right) C_A C_F T_F \Nf
+ \left( 10368 \zeta(3) - 560 \right) C_F^3 \right. \nonumber \\
&& \left. ~~~~~~~~~~+~ \left( 10368 \zeta(3) - 6824 \right) C_F^2 T_F \Nf
- 896 C_F T_F^2 \Nf^2 \right] a^3 ~+~ O(a^4) ~, \nonumber \\
\gamma^{W_2}_{12}(a) &=& -~ \frac{4}{3} C_F a ~+~ \frac{1}{27}
\left[ 56 C_F^2 - 188 C_A C_F + 64 C_F T_F \Nf \right] a^2 \nonumber \\
&& +~ \frac{1}{243} \left[ \left( 7776 \zeta(3) + 4264 \right) C_A C_F^2
- \left( 2592 \zeta(3) + 10460 \right) C_A^2 C_F \right. \nonumber \\
&& \left. ~~~~~~~~~~+~ \left( 5184 \zeta(3) + 3128 \right) C_A C_F T_F \Nf
- \left( 5184 \zeta(3) - 280 \right) C_F^3 \right. \nonumber \\
&& \left. ~~~~~~~~~~-~ \left( 5184 \zeta(3) - 3412 \right) C_F^2 T_F \Nf
+ 448 C_F T_F^2 \Nf^2 \right] a^3 ~+~ O(a^4) ~, \nonumber \\
\gamma^{W_2}_{22}(a) &=& O(a^4)
\label{anom2}
\end{eqnarray}
and 
\begin{eqnarray}
\gamma^{W_3}_{11}(a) &=& \frac{25}{6} C_F a ~+~ \frac{1}{432} \left[
8560 C_A C_F - 2035 C_F^2 - 3320 C_F T_F \Nf \right] a^2 \nonumber \\
&& +~ \frac{1}{15552} \left[ \left( 285120 \zeta(3) + 1778866 \right) C_A^2 C_F
- \left( 855360 \zeta(3) + 311213 \right) C_A C_F^2 \right. \nonumber \\
&& \left. ~~~~~~~~~~~~~-~ \left( 1036800 \zeta(3) + 497992 \right)
C_A C_F T_F \Nf + \left( 570240 \zeta(3) - 244505 \right) C_F^3
\right. \nonumber \\
&& \left. ~~~~~~~~~~~~~+~ \left( 1036800 \zeta(3) - 814508 \right)
C_F^2 T_F \Nf - 82208 C_F T_F^2 \Nf^2 \right] a^3 ~+~ O(a^4) \nonumber \\
\gamma^{W_3}_{12}(a) &=& -~ \frac{3}{2} C_F a ~+~ \frac{1}{144} \left[
81 C_F^2 - 848 C_A C_F + 424 C_F T_F \Nf \right] a^2 ~+~ O(a^3) ~, \nonumber \\
\gamma^{W_3}_{13}(a) &=& -~ \frac{1}{2} C_F a ~+~ \frac{1}{144} \left[
103 C_F^2 - 388 C_A C_F + 104 C_F T_F \Nf \right] a^2 ~+~ O(a^3) ~, 
\nonumber \\
\gamma^{W_3}_{22}(a) &=& \frac{8}{3} C_F a ~+~ \frac{1}{27} \left[ 376 C_A C_F
- 112 C_F^2 - 128 C_F T_F \Nf \right] a^2 \nonumber \\
&& +~ \frac{1}{243} \left[ \left( 5184 \zeta(3) + 20920 \right) C_A^2 C_F
- \left( 15552 \zeta(3) + 8528 \right) C_A C_F^2 \right. \nonumber \\
&& \left. ~~~~~~~~~~-~ \left( 10368 \zeta(3) + 6256 \right) C_A C_F T_F \Nf
+ \left( 10368 \zeta(3) - 560 \right) C_F^3 \right. \nonumber \\
&& \left. ~~~~~~~~~~+~ \left( 10368 \zeta(3) - 6824 \right) C_F^2 T_F \Nf
- 896 C_F T_F^2 \Nf^2 \right] a^3 ~+~ O(a^4) ~, \nonumber \\
\gamma^{W_3}_{23}(a) &=& -~ \frac{4}{3} C_F a ~+~ \frac{1}{27}
\left[ 56 C_F^2 - 188 C_A C_F + 64 C_F T_F \Nf \right] a^2 \nonumber \\
&& +~ \frac{1}{243} \left[ \left( 7776 \zeta(3) + 4264 \right) C_A C_F^2
- \left( 2592 \zeta(3) + 10460 \right) C_A^2 C_F \right. \nonumber \\
&& \left. ~~~~~~~~~~+~ \left( 5184 \zeta(3) + 3128 \right) C_A C_F T_F \Nf
- \left( 5184 \zeta(3) - 280 \right) C_F^3 \right. \nonumber \\
&& \left. ~~~~~~~~~~-~ \left( 5184 \zeta(3) - 3412 \right) C_F^2 T_F \Nf
+ 448 C_F T_F^2 \Nf^2 \right] a^3 ~+~ O(a^4) ~, \nonumber \\
\gamma^{W_3}_{33}(a) &=& O(a^4)
\label{anom3}
\end{eqnarray}
where $\zeta(z)$ is the Riemann zeta function, $a$~$=$~$g^2/(16\pi^2)$ and 
$\Nf$ is the number of massless quarks. The group Casimirs are defined by
\begin{equation}
T^a T^a ~=~ C_F ~~,~~ \mbox{Tr} \left( T^a T^b \right) ~=~ 
T_F \delta^{ab} ~~,~~ f^{acd} f^{bcd} ~=~ C_A \delta^{ab}
\end{equation}
where $T^a$ are the Lie group generators with structure functions $f^{abc}$ and
$1$~$\leq$~$a$~$\leq$~$\NA$ where $\NA$ is the dimension of the adjoint
representation.  
 
\begin{figure}[ht]
\hspace{6cm}
\epsfig{file=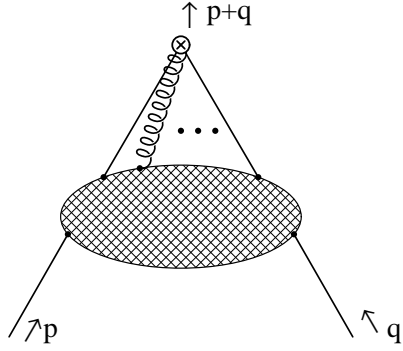,height=5cm}
\vspace{0.5cm}
\caption{Graphical illustration of the Green's function, $\left\langle \psi(p)
{\cal O}^i_{\mu_1 \ldots \mu_{n_i}} (-p-q) \bar{\psi}(q) \right\rangle$, used
to renormalize operators in the RI${}^\prime$/SMOM scheme.}
\end{figure}

We turn now to the set-up for the particular Green's function we are interested
in which is $\left\langle \psi(p) {\cal O}^i_{\mu_1 \ldots \mu_{n_i}} (-p-q) 
\bar{\psi}(q) \right\rangle$ and is illustrated in Figure $1$. The independent 
external momenta we use are $p$ and $q$ and are the momenta flowing into the 
external quark legs. Thus there is a momentum of $(p$~$+$~$q)$ flowing out 
through the operator insertion whose location is indicated by the circle 
containing a cross. In order to determine the renormalization constants for the
basic operators in the $\MSbar$ and RI${}^\prime$ schemes one chooses
$q$~$=$~$-$~$p$. However, for the RI${}^\prime$/SMOM scheme the two momenta are
left unconstrained. Instead to define the symmetric point of subtraction for 
the renormalization the square of the momenta satisfy, \cite{33,34,35}, 
\begin{equation}
p^2 ~=~ q^2 ~=~ ( p + q )^2 ~=~ -~ \mu^2 
\end{equation}
which imply
\begin{equation}
pq ~=~ \frac{1}{2} \mu^2 
\end{equation}
where $\mu$ is the renormalization scale introduced to ensure the coupling
constant is dimensionless in $d$-dimensions. Given this the Green's function is
decomposed into a basis of independent Lorentz tensors, ${\cal P}^i_{(k) \, 
\mu_1 \ldots \mu_{n_i} }(p,q)$, with associated amplitude, 
$\Sigma^{{\cal O}^i}_{(k)}(p,q)$, which is the value we will compute at the 
symmetric subtraction point, 
\begin{equation}
\left. \left\langle \psi(p) {\cal O}^i_{\mu_1 \ldots \mu_{n_i}}(-p-q) 
\bar{\psi}(q) \right\rangle \right|_{p^2 = q^2 = - \mu^2} ~=~ \sum_{k=1}^{n_i} 
{\cal P}^i_{(k) \, \mu_1 \ldots \mu_{n_i} }(p,q) \, 
\Sigma^{{\cal O}^i}_{(k)}(p,q) ~.
\end{equation}
Here the bracketed subscript $k$ labels the tensor of the basis and the 
superscript $i$ is the operator level label, (\ref{opdef}). The explicit 
tensors for each level are given in Appendix A together with the method which 
allows one to compute the amplitude itself via a projection onto the Green's 
function with free indices. The same tensor basis and projection is used for 
each level. The total number of projectors, $n_i$, is different for each level 
and recorded in Table $1$. It is worth noting that the basis of projection 
tensors which we use for each level is not unique. They are constructed from 
the basic momentum vectors, $\gamma$-matrices and metric tensors available, in 
such a way that each final tensor has the same symmetry structure as its 
associated operator insertion, as well as being traceless in $d$-dimensions. 
However, to construct all the tensors in each set of projectors we have 
isolated all the basic one loop tensor integrals within each computation. These
are then decomposed into their own tensor basis with their associated scalar 
integrals. For instance,
\begin{eqnarray}
\left. \int_k \frac{k_\mu k_\nu}{k^2 (k-p)^2 (k+q)^2} 
\right|_{p^2 \, = \, q^2 \, = \, - \mu^2} &=& \frac{4}{[d-2]} \left[
\eta_{\mu\nu} \left[ \frac{1}{4} I_1 + \frac{1}{3} I_2 + \frac{1}{6} I_3
+ \frac{1}{6} I_4 + \frac{1}{3} I_5 \right] \right. \nonumber \\
&& \left. ~~~~~~~~~~+~ \frac{p_\mu p_\nu}{\mu^2} \left[ \frac{1}{3} I_1 
+ \frac{4}{9} (d-1) I_2 + \frac{2}{9} (d-1) I_3 \right. \right. \nonumber \\
&& \left. \left. ~~~~~~~~~~~~~~~~~~~~~~+~ \frac{2}{9} (d-1) I_4 
+ \frac{1}{9} (d+2) I_5 \right] \right. \nonumber \\
&& \left. ~~~~~~~~~~+~ \frac{p_\mu q_\nu}{\mu^2} \left[ \frac{1}{6} I_1 
+ \frac{2}{9} (d-1) I_2 + \frac{1}{9} (4d-7) I_3 \right. \right. \nonumber \\
&& \left. \left. ~~~~~~~~~~~~~~~~~~~~~~+~ \frac{1}{9} (d-1) I_4
+ \frac{2}{9} (d-1) I_5  \right] \right. \nonumber \\
&& \left. ~~~~~~~~~~+~ \frac{q_\mu p_\nu}{\mu^2} \left[ \frac{1}{6} I_1 
+ \frac{2}{9} (d-1) I_2 + \frac{1}{9} (d-1) I_3 \right. \right. \nonumber \\
&& \left. \left. ~~~~~~~~~~~~~~~~~~~~~~+~ \frac{1}{9} (4d-7) I_4
+ \frac{2}{9} (d-1) I_5  \right] \right. \nonumber \\
&& \left. ~~~~~~~~~~+~ \frac{q_\mu q_\nu}{\mu^2} \left[ \frac{1}{3} I_1 
+ \frac{1}{9} (d+2) I_2 + \frac{2}{9} (d-1) I_3 \right. \right. \nonumber \\
&& \left. \left. ~~~~~~~~~~~~~~~~~~~~~~+~ \frac{2}{9} (d-1) I_4
+ \frac{4}{9} (d-1) I_5  \right] \right]
\label{ten2}
\end{eqnarray}
where $\int_k$~$=$~$\int \frac{d^dk}{(2\pi)^d}$ and
\begin{eqnarray}
I_1 &=& \left. \int_k \frac{1}{(k-p)^2 (k+q)^2} 
\right|_{p^2 \, = \, q^2 \, = \, - \mu^2} ~~,~~ 
I_2 ~=~ \left. \int_k \frac{(kp)^2}{k^2 (k-p)^2 (k+q)^2} 
\right|_{p^2 \, = \, q^2 \, = \, - \mu^2} ~, \nonumber \\
I_3 &=& I_4 ~=~ \left. \int_k \frac{kp \, kq}{k^2 (k-p)^2 (k+q)^2} 
\right|_{p^2 \, = \, q^2 \, = \, - \mu^2} ~~,~~ 
I_5 ~=~ \left. \int_k \frac{(kq)^2}{k^2 (k-p)^2 (k+q)^2} ~~~. \!\!\!\!\!\!\!\! 
\right|_{p^2 \, = \, q^2 \, = \, - \mu^2} 
\end{eqnarray}
are the basic scalar integrals. This together with the other relevant tensor 
integrals are substituted back into the main computation and the index 
contractions are performed. This ensures that no basis tensor is overlooked in 
the procedure of constructing the basis. Once the basis is established we 
determine the explicit projection of Appendix A. This is used to check the 
final results. Moreover, having found the tensor basis at one loop it can then 
be applied directly in a two loop calculation. 

{\begin{table}[ht]
\begin{center}
\begin{tabular}{|c||c|c|c|c|c|}
\hline
$i$ & $S$ & $V$ & $T$ & $W_2$ & $W_3$ \\
\hline
$n_{i}$ & $2$ & $6$ & $8$ & $10$ & $14$ \\
\hline
\end{tabular}
\end{center}
\begin{center}
{Table 1. Number of projectors for each operator insertion.}
\end{center}
\end{table}}

In addition, for the $\gamma$-algebra we use the generalized basis of 
$\gamma$-matrices which has been introduced in \cite{37,38,39} with more 
explicit details of their properties given in \cite{40,41}. Briefly, we define 
these new matrices, $\Gamma_{(n)}^{\mu_1 \ldots \mu_n}$, by
\begin{equation}
\Gamma_{(n)}^{\mu_1 \ldots \mu_n} ~=~ \gamma^{[\mu_1} \ldots \gamma^{\mu_n]}
\end{equation}
which is totally antisymmetric in the Lorentz indices for $n$~$\geq$~$1$ and
the square bracket notation includes the overall factor of $1/n!$. So, for
instance, $\Gamma_{(0)}$ is the unit matrix, $\Gamma_{(0)}$~$=$~$I$, and 
$\sigma^{\mu\nu}$~$=$~$\Gamma_{(2)}^{\mu\nu}$. Though we will invariably
retain $\gamma^\mu$ for $n$~$=$~$1$. One advantage of this basis is that
\begin{equation}
\mbox{tr} \left( \Gamma_{(m)}^{\mu_1 \ldots \mu_m} 
\Gamma_{(n)}^{\nu_1 \ldots \nu_n} \right) ~ \propto ~ \delta_{mn}
I^{\mu_1 \ldots \mu_m \nu_1 \ldots \nu_n} 
\label{gamtr}
\end{equation}
where $I^{\mu_1 \ldots \mu_m \nu_1 \ldots \nu_n}$ is the unit matrix in this
$\Gamma$-space, \cite{37,38,39,40,41}. As a result of this there is a
partitioning of the projection matrix into sectors with different 
$\Gamma_{(n)}$ as is evident in the examples in Appendix A. Two final points 
concerning the projectors are worth emphasising. First, our choice of basis 
tensors, ${\cal P}^i_{(k) \, \mu_1 \ldots \mu_{n_i} }(p,q)$, only has meaning 
strictly at the symmetric subtraction point. Away from this point there will be
a bigger basis set of tensors since then we would have $p^2$~$\neq$~$q^2$, 
$p^2$~$\neq$~$(p+q)^2$ and $q^2$~$\neq$~$(p+q)^2$ as is evident from the 
explicit forms given in Appendix A. Second, at the symmetric point the number 
of independent tensors in each case is clearly larger than that of the 
asymmetric forward subtraction point considered in the RI${}^\prime$ scheme. 
For instance, in the case of the scalar there are two tensors in the basis 
unlike the unique one of the RI${}^\prime$ scheme. In this case the additional
basis element is $\Gamma_{(2)}^{pq}$ where we use the convention that the 
appearance of a momentum vector in place of a Lorentz index indicates the 
contraction of that index with that momentum. Clearly this object vanishes if 
either momentum is zero or one is proportional to the other, corresponding to 
the RI${}^\prime$ scheme momentum configuration. As regards the tensor basis 
for each of the operators, the use of the generalized matrices 
$\Gamma_{(n)}^{\mu_1 \ldots \mu_n}$ is important in regarding the basis as 
complete. This is because they span spinor space in $d$-dimensions, 
\cite{37,39}, with $n$ being any positive integer. Therefore, it is natural to 
make use of them in dimensionally regularized computations. Though for the 
operators considered here $n$ never exceeds $4$. If there were more than two
independent momenta then obviously a larger value of $n$ would be necessary. As
$\Gamma_{(n)}^{\mu_1 \ldots \mu_n}$ clearly form the basis of the spinor vector
space of the tensor basis decomposition, the Lorentz vector space part of the
overall basis for each operator is then made complete by building Lorentz
tensors from combinations of elements of the set $\{ \eta^{\mu\nu}, p^\mu, 
q^\mu, \Gamma_{(n)}^{\mu_1 \ldots \mu_n} \}$. These have, of course, to be
consistent with the symmetries of the Lorentz indices of the operator inserted  
in the Green's function. From this it is clear that our basis is complete for 
each operator, as there is no room to build additional tensors from the basic
structures of the Lorentz and spinor vector spaces.

Given that there is more than one amplitude for each operator insertion, we
have to be careful in defining the renormalization constant in the 
RI${}^\prime$/SMOM scheme. For all the cases we consider here the ultraviolet
divergence resides in a subset of the amplitudes which in fact contains at 
least one element except for the special case of the vector current. For $V$
the RI${}^\prime$/SMOM scheme renormalization has to be treated separately. If,
for the moment, we denote this representative basis tensor by the label $0$ 
then we define the renormalization constant for the operator ${\cal O}$, 
$Z^{\mbox{\footnotesize{RI$^\prime$/SMOM}}}_{\cal O}$, by the condition
\begin{equation}
\left. \lim_{\epsilon\rightarrow 0} \left[ 
Z^{\mbox{\footnotesize{RI$^\prime$}}}_\psi 
Z^{\mbox{\footnotesize{RI$^\prime$/SMOM}}}_{\cal O} \Sigma^{\cal O}_{(0)}(p,q) 
\right] \right|_{p^2 \, = \, q^2 \, = \, - \mu^2} ~=~ 1 
\label{rissmomdef}
\end{equation} 
where $Z^{\mbox{\footnotesize{RI$^\prime$}}}_\psi$ is the quark wave function
renormalization constant in the RI${}^\prime$ scheme which is given in 
\cite{27,28}. The reason why the value in the original RI${}^\prime$ scheme is 
used has been discussed in \cite{33,34,35}. In determining the final
renormalization constant $Z^{\mbox{\footnotesize{RI$^\prime$/SMOM}}}_{\cal O}$,
we follow the procedure of \cite{42} for automatic Feynman diagram
computations. In other words we compute all diagrams in terms of their bare
quantities which here are essentially the coupling constant and the gauge 
parameter. Then the renormalized parameters are introduced by rescaling with 
the already determined coupling constant and gauge parameter renormalization
constants. Although the latter should be taken to be in the RI${}^\prime$ 
scheme to the one loop order we are working any scheme effect will not show up
until two loops. Whilst this is a standard procedure for introducing 
counterterms, the main issue here is that this rescaling from bare to 
renormalized quantities must also include the mixing of the operators.
Therefore, in constructing our amplitudes, which are recorded in sections
three and four, the matrices (\ref{mat2}) and (\ref{mat3}) have been included. 
In practical terms this means that the renormalization constants are found by 
first fixing those in the last row of each matrix. Then those in the next row 
are determined and repeated until the ultimate row is found. This is similar to
the method used in \cite{36} to deduce (\ref{anom2}) and (\ref{anom3}). The 
basic reason for this literal bottom up approach is that the counterterms to be
determined are intertwined due to the triangularity of the matrix and this is 
the systematic way to disentangle them. However, in defining the 
RI${}^\prime$/SMOM scheme here in the general terms indicated in 
(\ref{rissmomdef}) one has to be careful in any situation involving the vector 
current due to the underlying Slavnov-Taylor identity which affects this 
operator. We will discuss this caveat in more detail later as it arises not 
just in the case of the vector current itself but is embedded in each set $W_2$
and $W_3$.

Having concentrated on the general quantum field theoretic formalism that is
used, we now comment on the practicalities of the calculation. We use standard
tools for this. All the algebra is carried out with the symbolic manipulation
language {\sc Form}, \cite{43}. The three one loop Feynman diagrams are
generated in electronic form by the {\sc Qgraf} package, \cite{44}, with the 
output converted into {\sc Form} input notation. This procedure appends indices
and labels to all the fields. Various {\sc Form} modules were then used to
insert the basic Feynman rules for the propagators and vertices before those of
the particular operator of interest. The $n_i$ amplitudes were then projected 
out by the theory given in Appendix A for successive Green's functions. These
are first determined by constructing all the basic tensor integral 
decompositions akin to (\ref{ten2}). Once determined we construct the tensor
basis and repeat the computation by use of the projections given in Appendix A.
The final part is to actually substitute the explicit values of the master one 
loop scalar integrals. This may have required integration by parts but for the 
most part the resulting integrals are one loop with only two propagators such
as $I_1$. These bubbles are simple to determine. The remaining integral is
\begin{equation}
\left. \int_k \frac{1}{k^2 (k-p)^2 (k+q)^2} 
\right|_{p^2 \, = \, q^2 \, = \, - \mu^2} ~=~ \frac{9 s_2}{\mu^2} ~+~ 
O(\epsilon) 
\end{equation}
where $s_2$~$=$~$(2\sqrt{3}/9) \mbox{Cl}_2(2\pi/3)$ and $\mbox{Cl}_2(x)$ is the
Clausen function which was evaluated in \cite{45}. We use dimensional 
regularization in $d$~$=$~$4$~$-$~$2\epsilon$ dimensions. Given these 
ingredients we have been able to determine all the one loop amplitudes for the 
set of operators (\ref{opdef}) where we have repeated the calculation for $S$,
$V$ and $T$ as an elementary check on our programmes. We correctly reproduced 
all the one loop expressions given in \cite{33,34,35} for both anomalous 
dimensions and amplitudes. 

\sect{Quark currents.} 

Having concentrated on describing the background to the problem and the
methodology of the computations, we turn to the mundane task of recording the
explicit amplitudes. In this section we do this successively for the scalar,
vector and tensor currents. In \cite{33,34,35} the anomalous dimensions were
computed at one and two loops for the scalar and tensor cases in the 
RI${}^\prime$/SMOM schemes. In addition the one loop amplitudes in the $\MSbar$
scheme were given for each of the three operators in \cite{33}. In the appendix
we give the explicit mapping between our amplitudes, 
$\Sigma^{{\cal O}}_{(k)}(p,q)$, for the quark currents and those defined in 
\cite{33} in order to compare. Though we note that we are in full agreement 
with the results of \cite{33}. Since lattice computations will measure various 
directions and then extract estimates for the overall Green's function it seems
appropriate to provide the amplitudes for the $\MSbar$ scheme, as well as for 
RI${}^\prime$/SMOM, as ultimately the former is the reference scheme one will 
map to. Equally the explicit $\MSbar$ expressions will be useful for lattice 
groups who convert to $\MSbar$ first before comparing to the high energy 
expressions rather than work in the RI${}^\prime$/SMOM directly in order to do 
the matching. This seems appropriate since it will be noted later that there is
not a definitive RI${}^\prime$/SMOM scheme especially when operators have 
Lorentz indices such as the Wilson operators. Moreover, whilst providing 
results for both schemes may appear to introduce a degree of redundancy, 
because at one loop there is overlap in the actual expressions for the 
amplitudes between the schemes, this is not preserved at two loops, \cite{46}. 
Also for the Wilson operators the results in both schemes will be useful to 
ensure that we have preserved properties of the sets of operators in each level
between schemes. For the scalar current we have the $\MSbar$ result at one loop
\begin{eqnarray}
\left. \Sigma^{S}_{(1)}(p,q) \right|_{\MSbars} &=& -~ 1 ~+~ C_F \left[ 
\frac{27}{2} s_2 - 4 - 2 \alpha + \frac{9}{2} s_2 \alpha \right] a ~+~ 
O(a^2) ~, \nonumber \\
\left. \Sigma^{S}_{(2)}(p,q) \right|_{\MSbars} &=& C_F \left[ 3 s_2 \alpha 
- 3 s_2 \right] a ~+~ O(a^2) ~. 
\end{eqnarray}
To make contact with other work in this area, \cite{33}, we note the relations
\begin{equation}
C_0(1) ~=~ 9 s_2 ~=~ \frac{2}{3} \psi^\prime \left( \frac{1}{3} \right) ~-~
\left( \frac{2\pi}{3} \right)^2
\end{equation}
where $\psi(z)$ is the derivative of the logarithm of the Euler
$\Gamma$-function and $C_0(\omega)$ was the function used in \cite{33} to 
interpolate between the symmetric and exceptional momenta scheme choices. 
Whilst this case has already been analysed in \cite{33,34,35} we discuss it
here as it raises several issues with regard to defining the subsequent
RI${}^\prime$/SMOM scheme renormalization constants for the Wilson operators.
For instance, it would appear that the definition of the scheme can be given in
several ways. Given this form for the Green's function, one could either define
the RI${}^\prime$/SMOM scheme operator renormalization constant by absorbing 
the finite part of the $1$ direction. Alternatively one could take the 
projection of the Green's function by some tensor, such as that of the Born 
term, and then absorb whatever the finite part emerges into the renormalization
constant. For the scalar case this would actually give the same renormalization
constant because of (\ref{gamtr}). However, for other operators there appears 
to be a degree of freedom in how one can actually define the scheme. Though, of
course, one would retain the fundamental criterion of (\ref{rissmomdef}) to 
ensure there are no corrections beyond the leading order. As these methods are
equivalent here we note that we reproduce the scalar current results of
\cite{33} for the renormalization constant
\begin{equation}
Z^S ~=~ 1 ~+~ C_F \left[ -~ \frac{3}{\epsilon} - 4 + \frac{27}{2} s_2
- \alpha + \frac{9}{2} s_2 \alpha \right] a ~+~ O(a^2) 
\end{equation}
where we use the notation that all expressions will be in RI${}^\prime$/SMOM
unless explicitly indicated to be in $\MSbar$. The resulting amplitudes are
\begin{eqnarray}
\Sigma^{S}_{(1)}(p,q) &=& -~ 1 ~+~ O(a^2) ~, 
\nonumber \\
\Sigma^{S}_{(2)}(p,q) &=& C_F \left[ 3 s_2 \alpha - 3 s_2 \right] a ~+~ O(a^2) 
\end{eqnarray}
where channel $2$ is the same as the $\MSbar$ expression. However, this will 
not be the case at two loops, \cite{46}.

The situation for the vector current is more involved partly because of the
increase in amplitudes with one free Lorentz index. First, the $\MSbar$
amplitudes are 
\begin{eqnarray}
\left. \Sigma^{V}_{(1)}(p,q) \right|_{\MSbars} &=& -~ 1 ~+~ C_F \left[ 2 
- 3 s_2 - 2 \alpha + 6 s_2 \alpha \right] a ~+~ O(a^2) ~, \nonumber \\
\left. \Sigma^{V}_{(2)}(p,q) \right|_{\MSbars} &=& \left. \Sigma^{V}_{(5)}(p,q) 
\right|_{\MSbars} ~=~ C_F \left[ \frac{8}{3} - 6 s_2 - \frac{4}{3} \alpha
+ 6 s_2 \alpha \right] a ~+~ O(a^2) ~, \nonumber \\
\left. \Sigma^{V}_{(3)}(p,q) \right|_{\MSbars} &=& \left. \Sigma^{V}_{(4)}(p,q) 
\right|_{\MSbars} ~=~ C_F \left[ \frac{4}{3} - \frac{2}{3} \alpha 
+ 6 s_2 \alpha \right] a ~+~ O(a^2) ~, \nonumber \\
\left. \Sigma^{V}_{(6)}(p,q) \right|_{\MSbars} &=& -~ 6 C_F s_2 a ~+~ O(a^2) ~.
\label{vecrissmom}
\end{eqnarray}
Here the renormalization constant for the current is
\begin{equation}
\left. Z^V \right|_{\MSbars} ~=~ 1 ~+~ O(a^2) ~.
\label{zvms}
\end{equation}
This value derives from the fact that the vector current is a physical
operator and undergoes no renormalization to all orders in perturbation theory.
Moreover this is a consequence of the Slavnov-Taylor identity. Turning to the
RI${}^\prime$/SMOM situation, it is tempting to define the renormalization
constant in this case by either the projection by the Born tensor or by taking
the coefficient of the channel $1$ amplitude. This would lead to a 
renormalization constant in the RI${}^\prime$/SMOM scheme with a non-zero
finite part. However, this would be inconsistent with the Slavnov-Taylor
identity. In addition as the $\MSbar$ renormalization constant is unity to all
orders then this implies, \cite{47}, that the RI${}^\prime$/SMOM 
renormalization constant is already determined and equivalent to (\ref{zvms}). 
Therefore, we have for the vector current renormalization 
\begin{equation}
Z^V ~=~ 1 ~+~ O(a^2) ~. 
\end{equation}
In order to see that this is equivalent with the Slavnov-Taylor identity one
can contract the Green's function with the vector $(p+q)_\mu$ as this is the
momentum flowing through the inserted operator. This procedure corresponds to
the renormalization of the divergence of the current. In terms of 
(\ref{vecrissmom}) the combination of amplitudes in this contraction 
proportional to $\pslash$ is 
\begin{equation}
\Sigma^{V}_{(1)}(p,q) ~-~ \frac{1}{2} \Sigma^{V}_{(2)}(p,q) ~-~ \frac{1}{2}
\Sigma^{V}_{(5)}(p,q) ~=~ -~ 1 ~+~ O(a^2) ~. 
\label{vecstid}
\end{equation}
A different combination determines the coefficient for the piece involving
$\qslash$ but with the same result. The fact that there is no $O(a)$ correction 
is because we have renormalized the quark $2$-point functions in an 
RI${}^\prime$ scheme and the wave function renormalization constant defined in 
this way leaves the $2$-point function as unity to all orders. Hence the 
Slavnov-Taylor identity is satisfied. We have checked that this is also the 
situation in the $\MSbar$ case and, moreover, this has been extended to two 
loops in RI${}^\prime$/SMOM, \cite{46}, where it was verified that the 
Slavnov-Taylor identity was consistent in that case too in keeping with the 
general argument given in \cite{33}. Crucial to this analysis is knowledge of 
the full basis of tensors into which the Green's function can be written. 
Although the channel $6$ amplitude appears to be decoupled due to the 
$\Gamma_{(n)}$ basis we use, omitting it would lead to an inconsistent operator
renormalization. Finally, in considering the Slavnov-Taylor identity in this 
way it would have become clearer if a different combination of tensors was used
in the basis rather than the ones given in Appendix A. In other words we could 
have used a basis where all but one tensor in the basis vanished when 
contracted by $(p+q)_\mu$ whence the presence of the Slavnov-Taylor identity 
would have been explicit. Finally, we note the RI${}^\prime$/SMOM scheme 
amplitudes after renormalization are 
\begin{eqnarray}
\Sigma^{V}_{(1)}(p,q) &=& -~ 1 ~+~ C_F \left[ 2 - \alpha - 3 s_2 + 6 s_2 \alpha
\right] a  ~+~ O(a^2) ~, \nonumber \\
\Sigma^{V}_{(2)}(p,q) &=& \Sigma^{V}_{(5)}(p,q) ~=~ -~ C_F \left[ 6 s_2 
- \frac{8}{3} - 6 s_2 \alpha + \frac{4}{3} \alpha \right] a ~+~ O(a^2) ~, 
\nonumber \\
\Sigma^{V}_{(3)}(p,q) &=& \Sigma^{V}_{(4)}(p,q) ~=~ C_F \left[ \frac{4}{3} 
- \frac{2}{3} \alpha + 6 s_2 \alpha \right] a ~+~ O(a^2) ~, \nonumber \\
\Sigma^{V}_{(6)}(p,q) &=& -~ 6 C_F s_2 a ~+~ O(a^2) ~. 
\end{eqnarray}
As this particular Green's function is symmetric under swapping the two 
independent momenta, then two pairs of the amplitudes are equivalent. The 
different values for $\Sigma^{V}_{(1)}(p,q)$ in both schemes is due to the fact
that the finite parts of the quark wave function are not the same in each
scheme.

For the tensor case we provide the same information as the previous cases.
First, the $\MSbar$ amplitudes are 
\begin{eqnarray}
\left. \Sigma^{T}_{(1)}(p,q) \right|_{\MSbars} &=& -~ 1 ~+~ C_F \left[ 2 
- \frac{15}{2} s_2 - 2 \alpha + \frac{15}{2} s_2 \alpha \right] a ~+~ O(a^2) ~,
\nonumber \\
\left. \Sigma^{T}_{(2)}(p,q) \right|_{\MSbars} &=& C_F \left[ 9 s_2 
+ 3 s_2 \alpha \right] a ~+~ O(a^2) ~, \nonumber \\
\left. \Sigma^{T}_{(3)}(p,q) \right|_{\MSbars} &=& \left. \Sigma^{T}_{(6)}(p,q) 
\right|_{\MSbars} ~=~ C_F \left[ \frac{4}{3} - 6 s_2 - \frac{4}{3} \alpha
+ 6 s_2 \alpha \right] a ~+~ O(a^2) ~, \nonumber \\
\left. \Sigma^{T}_{(4)}(p,q) \right|_{\MSbars} &=& 
\left. \Sigma^{T}_{(5)}(p,q) \right|_{\MSbars} ~=~ C_F \left[ \frac{2}{3} 
- 3 s_2 - \frac{2}{3} \alpha + 3 s_2 \alpha \right] a ~+~ O(a^2) ~, 
\nonumber \\ 
\left. \Sigma^{T}_{(7)}(p,q) \right|_{\MSbars} &=& C_F \left[ \frac{8}{3} 
- 12 s_2 - \frac{8}{3} \alpha + 12 s_2 \alpha \right] a ~+~ O(a^2) ~, 
\nonumber \\
\left. \Sigma^{T}_{(8)}(p,q) \right|_{\MSbars} &=& -~ C_F \left[ 9 s_2 
+ 3 s_2 \alpha \right] a ~+~ O(a^2) ~. 
\end{eqnarray}
The earlier comments concerning how one actually defines the RI${}^\prime$/SMOM
scheme apply here. In \cite{33,35} the finite part of the tensor current
renormalization constant was defined by contracting the Green's function with
$\Gamma_{(2)}^{\mu\nu}$ and ensuring that the resulting expression had no $a$
dependence after renormalization. Following that with our amplitudes we find
exact agreement with the divergent and finite parts of the same renormalization
constant as, \cite{33}, 
\begin{equation}
Z^T ~=~ 1 ~+~ C_F \left[ \frac{1}{\epsilon} + \frac{4}{3}
- \frac{9}{2} s_2 - \frac{1}{3} \alpha + \frac{9}{2} s_2 \alpha 
\right] a ~+~ O(a^2) ~. 
\label{ztcons}
\end{equation}
As in the vector case with this projection to define the renormalization
constant the $\Gamma_{(0)}$ and $\Gamma_{(4)}$ sectors do not contribute but
are part of the full Green's function at the symmetric subtraction point.  
Indeed we have checked that the full expression for the two loop amplitude,
\cite{46}, leads to the same renormalization constant as \cite{35}. However, if
one were to follow an alternative way of defining the RI${}^\prime$/SMOM 
renormalization constant by merely using the coefficient of channel $1$ which 
is where the divergence resides then one would have another expression for 
$Z^T$ which is
\begin{equation}
\left. Z^T \right|_{\mbox{\footnotesize{alt RI${}^\prime$/SMOM}}} ~=~ 
1 ~+~ C_F \left[ \frac{1}{\epsilon} + 2 - \frac{15}{2} s_2 - \alpha 
+ \frac{15}{2} s_2 \alpha \right] a ~+~ O(a^2) ~. 
\end{equation}
This alternative definition is of course dependent on the choice of basis 
tensors. One feature of it is that the numerical value of the finite part in 
the Landau gauge is significantly smaller than the corresponding part of 
(\ref{ztcons}). Indeed with a judicious choice of the projection basis it might
be possible to have a rapidly converging series for the conversion function. 
However, that requires a higher loop analysis, \cite{46}. Returning to the 
original RI${}^\prime$/SMOM scheme definition of \cite{33,34,35} the amplitudes
are
\begin{eqnarray}
\Sigma^{T}_{(1)}(p,q) &=& -~ 1 ~+~ C_F \left[ \frac{2}{3} - 3 s_2 - \frac{2}{3}
\alpha + 3 s_2 \alpha \right] a ~+~ O(a^2) ~, \nonumber \\
\Sigma^{T}_{(2)}(p,q) &=& C_F \left[ 9 s_2 + 3 s_2 \alpha \right] a ~+~ O(a^2) 
\nonumber \\
\Sigma^{T}_{(3)}(p,q) &=& \Sigma^{T}_{(6)}(p,q) ~=~ C_F \left[ \frac{4}{3} 
- 6 s_2 - \frac{4}{3} \alpha + 6 s_2 \alpha \right] a ~+~ O(a^2) ~, 
\nonumber \\
\Sigma^{T}_{(4)}(p,q) &=& \Sigma^{T}_{(5)}(p,q) ~=~ C_F \left[ \frac{2}{3} 
- 3 s_2 - \frac{2}{3} \alpha + 3 s_2 \alpha \right] a ~+~ O(a^2) ~, 
\nonumber \\
\Sigma^{T}_{(7)}(p,q) &=& C_F \left[ \frac{8}{3} - 12 s_2 - \frac{8}{3} \alpha 
+ 12 s_2 \alpha \right] a ~+~ O(a^2) ~, \nonumber \\
\Sigma^{T}_{(8)}(p,q) &=& -~ C_F \left[ 9 s_2 + 3 s_2 \alpha \right] a ~+~ 
O(a^2) ~. 
\end{eqnarray}
Similar to the vector case there is a clear symmetry here because the Green's
function is symmetric under the interchange of $p$ and $q$ which is responsible
for the two pairs of equivalent amplitudes which is clear from the explicit
tensor definitions given in Appendix A.

\sect{Deep inelastic scattering operators.}

The situation for both moments of the flavour non-singlet twist-$2$ Wilson
operators is more involved due to the operator mixing issue as well as the
issue with the Slavnov-Taylor identity noted previously. First, the $\MSbar$
amplitudes for $n$~$=$~$2$ are 
\begin{eqnarray}
\left. \Sigma^{W_2}_{(1)}(p,q) \right|_{\MSbars} &=& C_F \left[ 2 s_2 
- \frac{11}{6} - \frac{1}{3} \alpha \right] a ~+~ O(a^2) ~, \nonumber \\
\left. \Sigma^{W_2}_{(2)}(p,q) \right|_{\MSbars} &=& -~ 1 ~+~ C_F \left[ 
\frac{23}{6} - 5 s_2 - \frac{5}{3} \alpha + 6 s_2 \alpha \right] a ~+~ 
O(a^2) ~, \nonumber \\
\left. \Sigma^{W_2}_{(3)}(p,q) \right|_{\MSbars} &=& C_F \left[ \frac{16}{27}
+ \frac{4}{3} s_2 - \frac{8}{9} \alpha + 4 s_2 \alpha \right] a ~+~ O(a^2) ~, 
\nonumber \\
\left. \Sigma^{W_2}_{(4)}(p,q) \right|_{\MSbars} &=& -~ C_F \left[ 
\frac{16}{3} s_2 - \frac{71}{27} - 8 s_2 \alpha + \frac{16}{9} \alpha 
\right] a ~+~ O(a^2) ~, \nonumber \\
\left. \Sigma^{W_2}_{(5)}(p,q) \right|_{\MSbars} &=& -~ C_F \left[ 
\frac{20}{3} s_2 - \frac{100}{27} - 16 s_2 \alpha + \frac{26}{9} \alpha 
\right] a ~+~ O(a^2) ~, \nonumber \\
\left. \Sigma^{W_2}_{(6)}(p,q) \right|_{\MSbars} &=& C_F \left[ 
\frac{20}{3} s_2 - \frac{28}{27} - 4 s_2 \alpha + \frac{14}{9} \alpha 
\right] a ~+~ O(a^2) ~, \nonumber \\
\left. \Sigma^{W_2}_{(7)}(p,q) \right|_{\MSbars} &=& -~ C_F \left[ 
\frac{2}{3} s_2 - \frac{37}{27} - 4 s_2 \alpha + \frac{2}{9} \alpha 
\right] a ~+~ O(a^2) ~, \nonumber \\
\left. \Sigma^{W_2}_{(8)}(p,q) \right|_{\MSbars} &=& -~ C_F \left[ 
\frac{40}{3} s_2 - \frac{128}{27} - 8 s_2 \alpha + \frac{16}{9} \alpha 
\right] a ~+~ O(a^2) ~, \nonumber \\
\left. \Sigma^{W_2}_{(9)}(p,q) \right|_{\MSbars} &=& -~ \frac{1}{3} a ~+~ 
O(a^2) ~, \nonumber \\
\left. \Sigma^{W_2}_{(10)}(p,q) \right|_{\MSbars} &=& C_F \left[ \frac{1}{3} 
- 6 s_2 \right]  a ~+~ O(a^2) ~, \nonumber \\
\left. \Sigma^{\partial W_2}_{(1)}(p,q) \right|_{\MSbars} &=& 
\left. \Sigma^{\partial W_2}_{(2)}(p,q) \right|_{\MSbars} ~=~ -~ 1 ~+~ 
C_F \left[ 2 - 3 s_2 - 2 \alpha + 6 s_2 \alpha \right] a ~+~ O(a^2) ~, 
\nonumber \\
\left. \Sigma^{\partial W_2}_{(3)}(p,q) \right|_{\MSbars} &=& 
\left. \Sigma^{\partial W_2}_{(8)}(p,q) \right|_{\MSbars} ~=~ -~ C_F \left[ 
12 s_2 - \frac{16}{3} - 12 s_2 \alpha + \frac{8}{3} \alpha
\right] a ~+~ O(a^2) ~, \nonumber \\
\left. \Sigma^{\partial W_2}_{(4)}(p,q) \right|_{\MSbars} &=& 
\left. \Sigma^{\partial W_2}_{(7)}(p,q) \right|_{\MSbars} ~=~ -~ C_F \left[ 
6 s_2 - 4 - 12 s_2 \alpha + 2 \alpha \right] a ~+~ O(a^2) ~, \nonumber \\
\left. \Sigma^{\partial W_2}_{(5)}(p,q) \right|_{\MSbars} &=&
\left. \Sigma^{\partial W_2}_{(6)}(p,q) \right|_{\MSbars} ~=~ -~
C_F \left[ \frac{4}{3} \alpha - \frac{8}{3} - 12 s_2 \alpha 
\right] a ~+~ O(a^2) ~,
\nonumber \\
\left. \Sigma^{\partial W_2}_{(9)}(p,q) \right|_{\MSbars} &=&
\left. \Sigma^{\partial W_2}_{(10)}(p,q) \right|_{\MSbars} ~=~ -~ 
6 C_F s_2 a ~+~ O(a^2) 
\end{eqnarray}
where the operator superscript label here corresponds to the row of the matrix
with that operator on the diagonal. Unlike the previous two cases there is now 
no symmetry for the Green's function itself when the original operator is 
inserted. This is because the covariant derivative in the operator only acts on
the quark and not the anti-quark. So swapping the external momenta in the 
Green's function is not a symmetric operation. By contrast for the associated 
total derivative operator this symmetry is still valid which is why there are 
equivalences between various pairs of amplitudes. Moreover, the actual 
expressions for the tensors labelled $3$ and $5$ are proportional to those 
respectively labelled $2$ and $3$ of the vector case. This is not unexpected
because the total derivative operator associated with this moment is 
effectively the vector current in disguise at the higher level. That the 
expressions are not precisely the same is due to the fact that one has an extra
Lorentz index present at this level so that the projection coefficient into the
basis will not have a completely parallel value. Although there is no unique 
basis for the projection tensors this agreement at one loop is a check on our 
calculational setup particularly since with another choice of tensors this 
proportionality could have been hidden. For the renormalization constants the 
relation between the divergences has already been noted at three loops in the 
$\MSbar$ scheme in \cite{36}.

Now that we have all the finite parts at the symmetric subtraction point we can
define the RI${}^\prime$/SMOM scheme renormalization constants. Similar to the 
vector current this requires care in the case of $\partial W_2$. This operator 
is the total derivative of the vector current whose renormalization constant is 
already determined in all schemes as it is a physical operator. However, only 
when one takes the contraction of the two free Lorentz indices of 
$\partial W_2$ does the divergence of the vector current emerge. Therefore, to
have a renormalization consistent with $Z^V$ we have to define $Z^{W_2}_{22}$
to be unity in the RI${}^\prime$/SMOM scheme. In the $\MSbar$ case, to 
contrast with (\ref{vecstid}), the appropriate combination of amplitudes for 
the piece proportional to $\pslash$ gives 
\begin{eqnarray}
&& -~ \frac{[d-2]}{d} \left. \Sigma^{\partial W_2}_{(1)}(p,q)
\right|_{\MSbars} ~-~  
\left. \Sigma^{\partial W_2}_{(2)}(p,q) \right|_{\MSbars} ~+~ 
\frac{[d-4]}{4d} \left. \Sigma^{\partial W_2}_{(3)}(p,q) \right|_{\MSbars}
\nonumber \\
&& +~ \frac{[d+2]}{2d} \left. \Sigma^{\partial W_2}_{(4)}(p,q) 
\right|_{\MSbars} ~+~ 
\frac{[d-4]}{4d} \left. \Sigma^{\partial W_2}_{(5)}(p,q) \right|_{\MSbars} ~=~ 
\frac{3}{2} ~+~ \frac{3}{2} C_F \alpha a ~+~ O(a^2)
\label{dw2stid}
\end{eqnarray}
whence it can be recognised that the Slavnov-Taylor identity is preserved. An
alternative choice gives the $\qslash$ contribution but it has the same result
as (\ref{dw2stid}). This combination is deduced by contracting with 
$(p+q)_\mu (p+q)_\nu$ to ensure each term in the definition of the operator 
involves the divergence of the vector current. However, for the other operator 
of level $W_2$ there is no underlying Slavnov-Taylor identity as the 
contraction of the two free Lorentz indices is not the divergence of a physical
current. Therefore, there is no constraint on how to define the remaining two 
elements of the $W_2$ renormalization constant matrix in RI${}^\prime$/SMOM. 
Indeed in some sense there is an infinite choice. However, for our purposes 
here we have chosen to define these renormalization constants by ensuring that 
there are no $O(a)$ corrections to channels $1$ and $2$ which both contain the 
divergences in $\epsilon$. The former contains the off-diagonal counterterm of 
the mixing matrix whilst the latter contains both counterterms from the first 
row of the matrix. Clearly one fixes the off-diagonal one first. Therefore, we 
have the $W_2$ matrix of renormalization constants
\begin{eqnarray}
Z^{W_2}_{11} &=& 1 ~+~ C_F \left[ \frac{8}{3\epsilon} + \frac{17}{3} - 7 s_2 
- \frac{1}{3} \alpha + 6 s_2 \alpha \right] a ~+~ O(a^2) ~, \nonumber \\
Z^{W_2}_{12} &=& C_F \left[ -~ \frac{4}{3\epsilon} - \frac{11}{6} + 2 s_2 
- \frac{1}{3} \alpha \right] a ~+~ O(a^2) ~, \nonumber \\
Z^{W_2}_{22} &=& 1 ~+~ O(a^2) 
\end{eqnarray}
where in defining $Z^{W_2}_{22}$ the right hand side of the analogous 
expression to (\ref{dw2stid}) we have ensured that there are no $O(a)$
corrections. Consequently, the RI${}^\prime$/SMOM amplitudes are
\begin{eqnarray}
\Sigma^{W_2}_{(1)}(p,q) &=& O(a^2) ~~~,~~~ 
\Sigma^{W_2}_{(2)}(p,q) ~=~ -~ 1 ~+~ O(a^2) ~, \nonumber \\
\Sigma^{W_2}_{(3)}(p,q) &=& C_F \left[ \frac{4}{3} s_2 + \frac{16}{27}
- \frac{8}{9} \alpha + 4 s_2 \alpha \right] a ~+~ O(a^2) ~, \nonumber \\
\Sigma^{W_2}_{(4)}(p,q) &=& -~ C_F \left[ \frac{16}{3} s_2 - \frac{71}{27} 
- 8 s_2 \alpha + \frac{16}{9} \alpha \right] a ~+~ O(a^2) ~, \nonumber \\
\Sigma^{W_2}_{(5)}(p,q) &=& -~ C_F \left[ \frac{20}{3} s_2 - \frac{100}{27} 
- 16 s_2 \alpha + \frac{26}{9} \alpha \right] a ~+~ O(a^2) ~, \nonumber \\
\Sigma^{W_2}_{(6)}(p,q) &=& C_F \left[ \frac{20}{3} s_2 - \frac{28}{27} 
- 4 s_2 \alpha + \frac{14}{9} \alpha \right] a ~+~ O(a^2) ~, \nonumber \\
\Sigma^{W_2}_{(7)}(p,q) &=& -~ C_F \left[ \frac{2}{3} s_2 - \frac{37}{27} 
- 4 s_2 \alpha + \frac{2}{9} \alpha \right] a ~+~ O(a^2) ~, \nonumber \\
\Sigma^{W_2}_{(8)}(p,q) &=& -~ C_F \left[ \frac{40}{3} s_2 - \frac{128}{27} 
- 8 s_2 \alpha + \frac{16}{9} \alpha \right] a ~+~ O(a^2) ~, \nonumber \\
\Sigma^{W_2}_{(9)}(p,q) &=& -~ \frac{1}{3} a ~+~ O(a^2) ~, \nonumber \\
\Sigma^{W_2}_{(10)}(p,q) &=& C_F \left[ \frac{1}{3} - 6 s_2 
\right] a ~+~ O(a^2) ~, \nonumber \\
\Sigma^{\partial W_2}_{(1)}(p,q) &=& \Sigma^{\partial W_2}_{(2)}(p,q) ~=~ 
-~ 1 ~+~ C_F \left[ 2 - 3 s_2 - \alpha + 6 s_2 \alpha \right] a ~+~ O(a^2) ~, 
\nonumber \\
\Sigma^{\partial W_2}_{(3)}(p,q) &=& \Sigma^{\partial W_2}_{(8)}(p,q) ~=~ -~
C_F \left[ 12 s_2 - \frac{16}{3} - 12 s_2 \alpha + \frac{8}{3} \alpha 
\right] a ~+~ O(a^2) ~, \nonumber \\
\Sigma^{\partial W_2}_{(4)}(p,q) &=& \Sigma^{\partial W_2}_{(7)}(p,q) ~=~ -~
C_F \left[ 6 s_2 - 4 - 12 s_2 \alpha + 2 \alpha \right] a ~+~ O(a^2) ~, 
\nonumber \\
\Sigma^{\partial W_2}_{(5)}(p,q) &=& \Sigma^{\partial W_2}_{(6)}(p,q) ~=~ -~
C_F \left[ \frac{4}{3} \alpha - \frac{8}{3} - 12 s_2 \alpha 
\right] a ~+~ O(a^2) ~, \nonumber \\
\Sigma^{\partial W_2}_{(9)}(p,q) &=& \Sigma^{\partial W_2}_{(10)}(p,q) ~=~ 
-~ 6 C_F s_2 a ~+~ O(a^2) ~. 
\end{eqnarray}
At this loop order the finite parts of a substantial number of the amplitudes
are the same as their $\MSbar$ counterparts. However, we do not expect this to
extend necessarily to two loops. Finally, we note that there is a relation 
between various amplitudes
\begin{equation}
\Sigma^{\partial W_2}_{(3)}(p,q) ~-~ 2 \Sigma^{\partial W_2}_{(4)}(p,q) ~+~ 
\Sigma^{\partial W_2}_{(5)}(p,q) ~=~ O(a^2)
\end{equation}  
which may still be valid at higher loop order.

For the next moment, $n$~$=$~$3$, the situation is of course more substantial 
since there are fourteen different basis tensors and three operators which mix.
First, we record the $\MSbar$ amplitudes to one loop are 
\begin{eqnarray}
\left. \Sigma^{W_3}_{(1)}(p,q) \right|_{\MSbars} &=& -~ C_F \left[ 
\frac{1}{108} + \frac{1}{3} s_2 + \frac{11}{54} \alpha - \frac{2}{3} s_2 \alpha
\right] a ~+~ O(a^2) ~, \nonumber \\ 
\left. \Sigma^{W_3}_{(2)}(p,q) \right|_{\MSbars} &=& -~ C_F \left[ 
\frac{35}{72} + \frac{5}{27} \alpha - \frac{1}{3} s_2 \alpha \right] a ~+~ 
O(a^2) ~, \nonumber \\
\left. \Sigma^{W_3}_{(3)}(p,q) \right|_{\MSbars} &=& -~ \frac{1}{3} ~+~ 
C_F \left[ \frac{203}{108} - \frac{8}{3} s_2 - \frac{35}{54} \alpha 
+ \frac{8}{3} s_2 \alpha \right] a ~+~ O(a^2) ~, \nonumber \\
\left. \Sigma^{W_3}_{(4)}(p,q) \right|_{\MSbars} &=& -~ C_F \left[ 2 s_2 
- \frac{7}{9} - \frac{26}{9} s_2 \alpha + \frac{58}{81} \alpha \right] a ~+~ 
O(a^2) ~, \nonumber \\
\left. \Sigma^{W_3}_{(5)}(p,q) \right|_{\MSbars} &=& -~ C_F \left[ 
\frac{8}{9} s_2 - \frac{89}{162} - \frac{22}{9} s_2 \alpha 
+ \frac{47}{81} \alpha \right] a ~+~ O(a^2) ~, \nonumber \\
\left. \Sigma^{W_3}_{(6)}(p,q) \right|_{\MSbars} &=& -~ C_F \left[
\frac{32}{9} s_2 - \frac{221}{162} - \frac{44}{9} s_2 \alpha
+ \frac{94}{81} \alpha \right] a ~+~ O(a^2) ~, \nonumber \\
\left. \Sigma^{W_3}_{(7)}(p,q) \right|_{\MSbars} &=& -~ C_F \left[ 
\frac{20}{3} s_2 - \frac{137}{54} - \frac{100}{9} s_2 \alpha 
+ \frac{194}{81} \alpha \right] a ~+~ O(a^2) ~, \nonumber \\
\left. \Sigma^{W_3}_{(8)}(p,q) \right|_{\MSbars} &=& -~ C_F \left[ 
\frac{14}{81} \alpha - \frac{1}{6} - \frac{10}{9} s_2 \alpha \right] a ~+~ 
O(a^2) ~, \nonumber \\
\left. \Sigma^{W_3}_{(9)}(p,q) \right|_{\MSbars} &=& C_F \left[ \frac{2}{9} s_2 
+ \frac{25}{162} + \frac{2}{9} s_2 \alpha + \frac{8}{81} \alpha \right] a ~+~ 
O(a^2) ~, \nonumber \\
\left. \Sigma^{W_3}_{(10)}(p,q) \right|_{\MSbars} &=& -~ C_F \left[ 
\frac{16}{9} s_2 - \frac{133}{162} - \frac{16}{9} s_2 \alpha 
+ \frac{17}{81} \alpha \right] a ~+~ O(a^2) ~, \nonumber \\
\left. \Sigma^{W_3}_{(11)}(p,q) \right|_{\MSbars} &=& -~ C_F \left[ 
\frac{28}{3} s_2 - \frac{77}{27} - \frac{44}{9} s_2 \alpha 
+ \frac{94}{81} \alpha \right] a ~+~ O(a^2) ~, \nonumber \\
\left. \Sigma^{W_3}_{(12)}(p,q) \right|_{\MSbars} &=& C_F \left[ \frac{1}{81}
- \frac{2}{9} s_2 \right] a ~+~ O(a^2) ~, \nonumber \\
\left. \Sigma^{W_3}_{(13)}(p,q) \right|_{\MSbars} &=& C_F \left[ \frac{2}{81}
- \frac{4}{9} s_2 \right] a ~+~ O(a^2) ~, \nonumber \\
\left. \Sigma^{W_3}_{(14)}(p,q) \right|_{\MSbars} &=& C_F \left[ \frac{19}{81}
- \frac{20}{9} s_2 \right] a ~+~ O(a^2) ~, \nonumber \\
\left. \Sigma^{\partial W_3}_{(1)}(p,q) \right|_{\MSbars} &=& -~ C_F \left[
\frac{11}{18} - \frac{2}{3} s_2 + \frac{1}{9} \alpha \right] a ~+~ O(a^2) ~,
\nonumber \\ 
\left. \Sigma^{\partial W_3}_{(2)}(p,q) \right|_{\MSbars} &=& 
-~ \frac{1}{6} ~+~ C_F \left[ \frac{1}{3} - \frac{1}{2} s_2 
- \frac{1}{3} \alpha + s_2 \alpha \right] a ~+~ O(a^2) ~, \nonumber \\
\left. \Sigma^{\partial W_3}_{(3)}(p,q) \right|_{\MSbars} &=& 
-~ \frac{1}{3} ~+~ C_F \left[ \frac{23}{18} - \frac{5}{3} s_2 
- \frac{5}{9} \alpha + 2 s_2 \alpha \right] a ~+~ O(a^2) ~, \nonumber \\
\left. \Sigma^{\partial W_3}_{(4)}(p,q) \right|_{\MSbars} &=& 
-~ C_F \left[ \frac{4}{9} \alpha - \frac{2}{3} s_2 - \frac{8}{27} 
- 2 s_2 \alpha \right] a ~+~ O(a^2) ~, \nonumber \\
\left. \Sigma^{\partial W_3}_{(5)}(p,q) \right|_{\MSbars} &=& -~ C_F \left[ 
\frac{14}{9} s_2 - \frac{79}{81} - \frac{10}{3} s_2 \alpha 
+ \frac{20}{27} \alpha \right] a ~+~ O(a^2) ~, \nonumber \\
\left. \Sigma^{\partial W_3}_{(6)}(p,q) \right|_{\MSbars} &=& -~ C_F \left[ 
\frac{26}{9} s_2 - \frac{121}{81} - \frac{16}{3} s_2 \alpha 
+ \frac{29}{27} \alpha \right] a ~+~ O(a^2) ~, \nonumber \\
\left. \Sigma^{\partial W_3}_{(7)}(p,q) \right|_{\MSbars} &=& -~ C_F \left[ 
\frac{10}{3} s_2 - \frac{50}{27} - 8 s_2 \alpha + \frac{13}{9} \alpha 
\right] a ~+~ O(a^2) ~, \nonumber \\
\left. \Sigma^{\partial W_3}_{(8)}(p,q) \right|_{\MSbars} &=& C_F \left[ 
\frac{10}{3} s_2 - \frac{14}{27} - 2 s_2 \alpha + \frac{7}{9} \alpha 
\right] a ~+~ O(a^2) ~, \nonumber \\
\left. \Sigma^{\partial W_3}_{(9)}(p,q) \right|_{\MSbars} &=& C_F \left[ 
\frac{8}{9} s_2 + \frac{23}{81} + \frac{2}{3} s_2 \alpha 
+ \frac{5}{27} \alpha \right] a ~+~ O(a^2) ~, \nonumber \\
\left. \Sigma^{\partial W_3}_{(10)}(p,q) \right|_{\MSbars} &=& -~ C_F \left[ 
\frac{22}{9} s_2 - \frac{101}{81} - \frac{8}{3} s_2 \alpha 
+ \frac{10}{27} \alpha \right] a ~+~ O(a^2) ~, \nonumber \\
\left. \Sigma^{\partial W_3}_{(11)}(p,q) \right|_{\MSbars} &=& -~ C_F \left[ 
\frac{20}{3} s_2 - \frac{64}{27} - 4 s_2 \alpha + \frac{8}{9} \alpha 
\right] a ~+~ O(a^2) ~, \nonumber \\
\left. \Sigma^{\partial W_3}_{(12)}(p,q) \right|_{\MSbars} &=& 
-~ \frac{1}{9} a ~+~ O(a^2) ~~,~~ 
\left. \Sigma^{\partial W_3}_{(13)}(p,q) \right|_{\MSbars} ~=~ -~ C_F s_2 a ~+~ 
O(a^2) ~, \nonumber \\ 
\left. \Sigma^{\partial W_3}_{(14)}(p,q) \right|_{\MSbars} &=& C_F \left[ 
\frac{1}{9} - 2 s_2 \right] a ~+~ O(a^2) ~, \nonumber \\
\left. \Sigma^{\partial\partial W_3}_{(1)}(p,q) \right|_{\MSbars} &=&
-~ \frac{1}{3} ~+~ C_F \left[ \frac{2}{3} - s_2 - \frac{2}{3} \alpha 
+ 2 s_2 \alpha \right] a ~+~ O(a^2) ~, \nonumber \\
\left. \Sigma^{\partial\partial W_3}_{(2)}(p,q) \right|_{\MSbars} &=&
-~ \frac{1}{3} ~+~ C_F \left[ \frac{2}{3} - s_2 - \frac{2}{3} \alpha 
+ 2 s_2 \alpha \right] a ~+~ O(a^2) ~, \nonumber \\
\left. \Sigma^{\partial\partial W_3}_{(3)}(p,q) \right|_{\MSbars} &=& 
-~ \frac{1}{3} ~+~ C_F \left[ \frac{2}{3} - s_2 - \frac{2}{3} \alpha 
+ 2 s_2 \alpha \right] a ~+~ O(a^2) ~, \nonumber \\
\left. \Sigma^{\partial\partial W_3}_{(4)}(p,q) \right|_{\MSbars} &=&
\left. \Sigma^{\partial\partial W_3}_{(11)}(p,q) \right|_{\MSbars} ~=~ 
-~ C_F \left[ 6 s_2 - \frac{8}{3} - 6 s_2 \alpha + \frac{4}{3} \alpha 
\right] a ~+~ O(a^2) ~, \nonumber \\
\left. \Sigma^{\partial\partial W_3}_{(5)}(p,q) \right|_{\MSbars} &=&
\left. \Sigma^{\partial\partial W_3}_{(10)}(p,q) \right|_{\MSbars} ~=~ 
-~ C_F \left[ 4 s_2 - \frac{20}{9} - 6 s_2 \alpha + \frac{10}{9} \alpha 
\right] a ~+~ O(a^2) ~, \nonumber \\
\left. \Sigma^{\partial\partial W_3}_{(6)}(p,q) \right|_{\MSbars} &=&
\left. \Sigma^{\partial\partial W_3}_{(9)}(p,q) \right|_{\MSbars} ~=~ -~ C_F 
\left[ 2 s_2 - \frac{16}{9} - 6 s_2 \alpha + \frac{8}{9} \alpha \right] a ~+~ 
O(a^2) ~, \nonumber \\
\left. \Sigma^{\partial\partial W_3}_{(7)}(p,q) \right|_{\MSbars} &=&
\left. \Sigma^{\partial\partial W_3}_{(8)}(p,q) \right|_{\MSbars} ~=~ -~ C_F 
\left[ \frac{2}{3} \alpha - \frac{4}{3} - 6 s_2 \alpha \right] a ~+~ O(a^2) ~,
\nonumber \\
\left. \Sigma^{\partial\partial W_3}_{(12)}(p,q) \right|_{\MSbars} &=&
\left. \Sigma^{\partial\partial W_3}_{(13)}(p,q) \right|_{\MSbars} ~=~
\left. \Sigma^{\partial\partial W_3}_{(14)}(p,q) \right|_{\MSbars} ~=~ 
-~ 2 C_F s_2 a ~+~ O(a^2) ~.
\end{eqnarray}
Whilst there are more amplitudes some features are similar to level $W_2$ such
as equivalences between certain amplitudes for $\partial \partial W_3$ and a
proportionality with various amplitudes in $V$ and $\partial W_2$. As this is a
level higher there is also a proportionality of certain $\partial W_3$
amplitudes with $W_2$. For instance, the channels $2$, $3$ and $4$ respectively
of $V$, $\partial W_2$ and $\partial \partial W_3$ are proportional as well as 
the respective channels $3$, $5$ and $7$. The same is true for $W_2$ and
$\partial W_3$ for the two respective pairs of channel $3$ and $4$ as well as
$5$ and $7$. This is of course not unexpected since the operators are
successive total derivatives of the lower level one. However, the 
renormalization of $\partial \partial W_3$ in the RI${}^\prime$/SMOM scheme is 
already predetermined for the same reasons as $\partial W_2$. Though in the 
case of $\partial \partial W_3$ one has to contract with $(p+q)_\mu (p+q)_\nu 
(p+q)_\sigma$ to produce the relevant divergence of the vector current. Using 
$(p+q)_\mu \eta_{\nu\sigma}$ as an alternative nullifies the operator as it is 
constucted to be traceless and gives nothing non-trivial. With the former 
contraction this produces the relation between the amplitudes involving 
$\pslash$ is
\begin{eqnarray}
&& \frac{3[d-6]}{4[d+2]} \Sigma^{\partial\partial W_3}_{(1)}(p,q) ~+~ 
\frac{3}{2} \Sigma^{\partial\partial W_3}_{(2)}(p,q) ~+~ 
\frac{3[d-4]}{4[d+2]} \Sigma^{\partial\partial W_3}_{(3)}(p,q) ~-~ 
\frac{[d-10]}{8[d+2]} \Sigma^{\partial\partial W_3}_{(4)}(p,q) \nonumber \\
&& -~ \frac{3}{8} \Sigma^{\partial\partial W_3}_{(5)}(p,q) ~-~ 
\frac{3}{8} \Sigma^{\partial\partial W_3}_{(6)}(p,q) ~-~ 
\frac{[d-10]}{8[d+2]} \Sigma^{\partial\partial W_3}_{(7)}(p,q) ~=~ 
-~ \frac{1}{2} ~+~ O(a^2) ~. 
\end{eqnarray}
An analogous relation produces the piece involving $\qslash$ which also has no
$O(a)$ piece consistent with the RI${}^\prime$ choice for the quark wave
function renormalization. Clearly this combination ensures consistency with the
Slavnov-Taylor identities. 
 
Again the choice of how to determine the remaining renormalization constants
within the RI${}^\prime$/SMOM scheme ethos is relatively free. The point of 
view we take to do this is to build on the previous level. Though there is no
reason why one should necessarily make this the definitive choice. By building
on the previous level we parallel the way we ensured that the hidden vector
current renormalization constant was determined. As $\partial W_3$ is the total
derivative of $W_2$ we choose that level of the renormalization constant mixing
matrix to be the same renormalization constants as the first row of the $W_2$
matrix. This leaves the first row of $W_3$ and we then define this by ensuring
that after renormalization there are no $O(a)$ corrections in the tensor basis
which originally had a divergence in $\epsilon$. As there is mixing within the
counterterms we have to determine these in sequence which was the same as the
$\MSbar$ scheme, \cite{36}. Therefore we have 
\begin{eqnarray}
Z^{W_3}_{11} &=& 1 ~+~ C_F \left[ \frac{25}{6\epsilon} + \frac{307}{36}
- 9 s_2 - \frac{4}{9} \alpha + 8 s_2 \alpha \right] a ~+~ O(a^2) ~, 
\nonumber \\
Z^{W_3}_{12} &=& C_F \left[ -~ \frac{3}{2\epsilon} - \frac{103}{36}
+ 2 s_2 + \frac{1}{9} \alpha - 2 s_2 \alpha \right] a ~+~ O(a^2) ~, 
\nonumber \\
Z^{W_3}_{13} &=& C_F \left[ -~ \frac{1}{2\epsilon} - \frac{1}{36}
- s_2 - \frac{11}{18} \alpha + 2 s_2 \alpha \right] a ~+~ O(a^2) ~, 
\nonumber \\
Z^{W_3}_{22} &=& 1 ~+~ C_F \left[ \frac{8}{3\epsilon} + \frac{17}{3}
- 7 s_2 - \frac{1}{3} \alpha + 6 s_2 \alpha \right] a ~+~ O(a^2) ~, 
\nonumber \\
Z^{W_3}_{23} &=& C_F \left[ -~ \frac{4}{3\epsilon} - \frac{11}{6}
+ 2 s_2 - \frac{1}{3} \alpha \right] a ~+~ O(a^2) ~, \nonumber \\
Z^{W_3}_{33} &=& 1 ~+~ O(a^2) ~. 
\end{eqnarray}
Once these have been determined the RI${}^\prime$/SMOM scheme amplitudes are
\begin{eqnarray}
\Sigma^{W_3}_{(1)}(p,q) &=& O(a^2) ~~~,~~~ \Sigma^{W_3}_{(2)}(p,q) ~=~ 
O(a^2) ~, \nonumber \\
\Sigma^{W_3}_{(3)}(p,q) &=& -~ \frac{1}{3} ~+~ O(a^2) ~, \nonumber \\
\Sigma^{W_3}_{(4)}(p,q) &=& -~ C_F \left[ 2 s_2 - \frac{7}{9} 
- \frac{26}{9} s_2 \alpha + \frac{58}{81} \alpha \right] a ~+~ O(a^2) ~, 
\nonumber \\
\Sigma^{W_3}_{(5)}(p,q) &=& -~ C_F \left[ \frac{8}{9} s_2 - \frac{89}{162} 
- \frac{22}{9} s_2 \alpha + \frac{47}{81} \alpha \right] a ~+~ O(a^2) ~, 
\nonumber \\
\Sigma^{W_3}_{(6)}(p,q) &=& -~ C_F \left[ \frac{32}{9} s_2 - \frac{221}{162} 
- \frac{44}{9} s_2 \alpha + \frac{94}{81} \alpha \right] a ~+~ O(a^2) ~,
\nonumber \\
\Sigma^{W_3}_{(7)}(p,q) &=& -~ C_F \left[ \frac{20}{3} s_2 - \frac{137}{54} 
- \frac{100}{9} s_2 \alpha + \frac{194}{81} \alpha \right] a ~+~ O(a^2) ~,
\nonumber \\
\Sigma^{W_3}_{(8)}(p,q) &=& -~ C_F \left[ \frac{14}{81} \alpha - \frac{1}{6} 
- \frac{10}{9} s_2 \alpha \right] a ~+~ O(a^2) ~, \nonumber \\
\Sigma^{W_3}_{(9)}(p,q) &=& C_F \left[ \frac{2}{9} s_2 + \frac{25}{162} 
+ \frac{2}{9} s_2 \alpha + \frac{8}{81} \alpha \right] a ~+~ O(a^2) ~,
\nonumber \\
\Sigma^{W_3}_{(10)}(p,q) &=& -~ C_F \left[ \frac{16}{9} s_2 - \frac{133}{162} 
- \frac{16}{9} s_2 \alpha + \frac{17}{81} \alpha \right] a ~+~ O(a^2) ~,
\nonumber \\
\Sigma^{W_3}_{(11)}(p,q) &=& -~ C_F \left[ \frac{28}{3} s_2 - \frac{77}{27} 
- \frac{44}{9} s_2 \alpha + \frac{94}{81} \alpha \right] a ~+~ O(a^2) ~,
\nonumber \\
\Sigma^{W_3}_{(12)}(p,q) &=& -~ C_F \left[ \frac{2}{9} s_2 - \frac{4}{81} 
\right] a ~+~ O(a^2) ~, \nonumber \\
\Sigma^{W_3}_{(13)}(p,q) &=& -~ C_F \left[ \frac{4}{9} s_2 - \frac{2}{81} 
\right] a ~+~ O(a^2) ~, \nonumber \\
\Sigma^{W_3}_{(14)}(p,q) &=& -~ C_F \left[ \frac{20}{9} s_2 - \frac{19}{81} 
\right] a ~+~ O(a^2) ~, \nonumber \\
\Sigma^{\partial W_3}_{(1)}(p,q) &=& O(a^2) ~~~,~~~ 
\Sigma^{\partial W_3}_{(2)}(p,q) ~=~ -~ \frac{1}{6} ~+~ O(a^2) ~, \nonumber \\
\Sigma^{\partial W_3}_{(3)}(p,q) &=& -~ \frac{1}{3} ~+~ O(a^2) ~, \nonumber \\
\Sigma^{\partial W_3}_{(4)}(p,q) &=& -~ C_F \left[ \frac{4}{9} \alpha 
- \frac{2}{3} s_2 - \frac{8}{27} - 2 s_2 \alpha \right] a ~+~ O(a^2) ~,
\nonumber \\
\Sigma^{\partial W_3}_{(5)}(p,q) &=& -~ C_F \left[ \frac{14}{9} s_2 
- \frac{79}{81} - \frac{10}{3} s_2 \alpha + \frac{20}{27} \alpha \right] a ~+~ 
O(a^2) ~, \nonumber \\
\Sigma^{\partial W_3}_{(6)}(p,q) &=& -~ C_F \left[ \frac{26}{9} s_2 
- \frac{121}{81} - \frac{16}{3} s_2 \alpha + \frac{29}{27} \alpha 
\right] a ~+~ O(a^2) ~, \nonumber \\
\Sigma^{\partial W_3}_{(7)}(p,q) &=& -~ C_F \left[ \frac{10}{3} s_2 
- \frac{50}{27} - 8 s_2 \alpha + \frac{13}{9} \alpha \right] a ~+~ O(a^2) ~, 
\nonumber \\
\Sigma^{\partial W_3}_{(8)}(p,q) &=& C_F \left[ \frac{10}{3} s_2 
- \frac{14}{27} - 2 s_2 \alpha + \frac{7}{9} \alpha \right] a ~+~ O(a^2) ~, 
\nonumber \\
\Sigma^{\partial W_3}_{(9)}(p,q) &=& C_F \left[ \frac{8}{9} s_2 + \frac{23}{81} 
+ \frac{2}{3} s_2 \alpha + \frac{5}{27} \alpha \right] a ~+~ O(a^2) ~,
\nonumber \\
\Sigma^{\partial W_3}_{(10)}(p,q) &=& -~ C_F \left[ \frac{22}{9} s_2 
- \frac{101}{81} - \frac{8}{3} s_2 \alpha + \frac{10}{27} \alpha 
\right] a ~+~ O(a^2) ~, \nonumber \\
\Sigma^{\partial W_3}_{(11)}(p,q) &=& -~ C_F \left[ \frac{20}{3} s_2 
- \frac{64}{27} - 4 s_2 \alpha + \frac{8}{9} \alpha \right] a ~+~ O(a^2) ~,
\nonumber \\
\Sigma^{\partial W_3}_{(12)}(p,q) &=& -~ \frac{1}{9} a ~+~ O(a^2) ~~,~~
\Sigma^{\partial W_3}_{(13)}(p,q) ~=~ -~ C_F s_2 a ~+~ O(a^2) ~, \nonumber \\
\Sigma^{\partial W_3}_{(14)}(p,q) &=& C_F \left[ \frac{1}{9} - 2 s_2 
\right] a ~+~ O(a^2) ~, \nonumber \\
\Sigma^{\partial\partial W_3}_{(1)}(p,q) &=&
\Sigma^{\partial\partial W_3}_{(2)}(p,q) \nonumber \\
&=& \Sigma^{\partial\partial W_3}_{(3)}(p,q) ~=~ -~ \frac{1}{3} ~+~ C_F \left[
\frac{2}{3} - s_2 - \frac{1}{3} \alpha + 2 s_2 \alpha \right] a ~+~ O(a^2) ~,
\nonumber \\
\Sigma^{\partial\partial W_3}_{(4)}(p,q) &=&
\Sigma^{\partial\partial W_3}_{(11)}(p,q) ~=~ -~ C_F \left[ 6 s_2 - \frac{8}{3} 
- 6 s_2 \alpha + \frac{4}{3} \alpha \right] a ~+~ O(a^2) ~, \nonumber \\
\Sigma^{\partial\partial W_3}_{(5)}(p,q) &=&
\Sigma^{\partial\partial W_3}_{(10)}(p,q) ~=~ -~ C_F \left[ 4 s_2 
- \frac{20}{9} - 6 s_2 \alpha + \frac{10}{9} \alpha \right] a ~+~ O(a^2) ~, 
\nonumber \\
\Sigma^{\partial\partial W_3}_{(6)}(p,q) &=&
\Sigma^{\partial\partial W_3}_{(9)}(p,q) ~=~ -~ C_F \left[ 2 s_2 - \frac{16}{9} 
- 6 s_2 \alpha + \frac{8}{9} \alpha \right] a ~+~ O(a^2) ~, \nonumber \\
\Sigma^{\partial\partial W_3}_{(7)}(p,q) &=&
\Sigma^{\partial\partial W_3}_{(8)}(p,q) ~=~ -~ C_F \left[ \frac{2}{3} \alpha 
- \frac{4}{3} - 6 s_2 \alpha \right] a ~+~ O(a^2) ~, \nonumber \\
\Sigma^{\partial\partial W_3}_{(12)}(p,q) &=&
\Sigma^{\partial\partial W_3}_{(13)}(p,q) ~=~
\Sigma^{\partial\partial W_3}_{(14)}(p,q) ~=~ -~ 2 C_F s_2 a ~+~ O(a^2) ~.
\end{eqnarray}
As with the lower moment there are similar relations to one loop. First, there 
is an obvious parallel with the vector and total derivative operator for 
$n$~$=$~$2$ with the double total derivative of the third moment which was 
noted in \cite{36}. In addition there are cross connections with the central 
row of the mixing matrix. Specifically, the amplitudes labelled $6$ and $8$ of 
$W_2$ are proportional to $8$ and $11$ of $\partial W_3$ respectively for the 
same reasons as before. Further, there is another apparent relation within the 
level since 
\begin{equation}
\Sigma^{\partial\partial W_3}_{(5)}(p,q) ~-~ 
2 \Sigma^{\partial\partial W_3}_{(6)}(p,q) ~+~ 
\Sigma^{\partial\partial W_3}_{(7)}(p,q) ~=~ O(a^2)
\end{equation}  
in addition to the reflection of the one already noted for moment $n$~$=$~$2$.
As an aside we note that in highlighting relations between amplitudes, we have 
concentrated on those which are true for all values of $\alpha$. There appears 
to be other relations which are only valid to one loop in the Landau gauge and 
since the case for regarding these as significant is diminished because of the 
apparent gauge dependence we do not draw attention to them.

As our computations have been at one loop the conversion functions that are
used to convert from the $\MSbar$ scheme to the RI${}^\prime$/SMOM scheme are
in effect the finite parts of the renormalization constants themselves. See, 
for example, \cite{47}. Therefore, in order to appreciate the magnitude of the 
one loop contribution it is a straightforward exercise to evaluate the finite 
parts numerically. It turns out that the correction increases in value with the
operator moment, in parallel, for example, with the numerical value of the one 
loop anomalous dimension coefficient. If we take the Landau gauge case and
compare the finite parts, the moments $n$~$=$~$2$ and $3$ are respectively, 
$3.8436$ and $6.1839$ where we use the $(11)$ element for the latter two or 
equivalently the diagonal of the $W_3$ matrix. This is with ignoring the common
colour factor, $C_F$, and coupling constant, $a$. Therefore, for higher moments
it would seem that one would require higher loop corrections in order to have a
more reliable convergence of the perturbative series for the conversion 
functions. Comparing with the RI${}^\prime$ scheme results of \cite{28,29,30} 
the respective numbers are $3.4444$ and $5.9444$. Unlike the situation with the
mass operator the RI${}^\prime$/SMOM scheme corrections for $W_2$ and $W_3$ are
slightly larger. Though in some sense this is not a fair comparison because of 
the mixing. The RI${}^\prime$ scheme is based on a specific momentum routing in
the Green's function which is blind to the off-diagonal matrix element of the 
renormalization constant matrix. Equally the convergence of the 
RI${}^\prime$/SMOM series could be improved by choosing a different combination
of the finite parts to define the actual renormalization constant since there 
appears to be a large degree of freedom in doing this. Only a higher order 
computation could give insight into this, \cite{46}. 

\sect{Discussion.}

We have determined the renormalization constants at one loop for the twist-$2$
flavour non-singlet operators with moments $n$~$=$~$2$ and $3$ in the
RI${}^\prime$/SMOM scheme and have taken account of operator mixing. In the 
Landau gauge the correction increases in value with the operator moment. So it 
would appear that the series for the conversion function will converge slower 
at high moment. Of course, a one loop computation is not sufficient to make a 
definitive statement since the higher loop corrections may produce an 
improvement. This is currently under investigation, \cite{46}. Equally it 
would be useful to see how the lattice measurements in this symmetric scheme 
for the deep inelastic scattering operators compares with the same analysis in 
the earlier RI${}^\prime$ scheme. Though it may be the case that not all the 
various tensor projections would give a clear accurate numerical signal. 
Finally, whilst the RI${}^\prime$/SMOM scheme is designed in order to 
circumvent infrared problems, one is still restricted to the Landau gauge due 
to the fact that the Green's function of Figure 1 is a gauge dependent 
quantity. Whilst this is not a problem for the high energy r\'{e}gime, it may 
be the case that in the infrared there are additional complications with Gribov
copies and hence the definitive measurement of operator matrix elements.  

\vspace{1cm}
\noindent
{\bf Acknowledgement.} The author thanks Dr. P.E.L. Rakow and Prof. C.T.C.
Sachrajda for useful discussions and especially the former for a careful
reading of the manuscript.

\appendix

\sect{Projectors.}

In this appendix we record in succession the basis of projection tensors we
have used in each of the various cases. These were denoted earlier by the
general notation ${\cal P}^i_{(k) \, \mu_1 \ldots \mu_{n_i} }(p,q)$. The 
matrix, ${\cal M}^i_{kl}$, in the second part of each subsection contains the 
coefficients associated with each basis projection tensor, which allows one to 
project out the amplitudes $\Sigma^{{\cal O}^i}_{(k)}(p,q)$. It is constructed 
by first determining the matrix
\begin{equation}
{\cal N}^i_{kl} ~=~ \left. {\cal P}^i_{(k) \, \mu_1 \ldots \mu_{n_i}}(p,q)
{\cal P}^{i ~\, \mu_1 \ldots \mu_{n_i}}_{(l)}(p,q) \right|_{p^2=q^2=-\mu^2} 
\end{equation}
where there is no summation over the label $i$ and $k$ and $l$ index the
projection tensors. The matrix ${\cal N}^i_{kl}$ is symmetric in $k$ and $l$. 
Given that the elements are contractions of Lorentz tensors in $d$-dimensions 
then the elements of ${\cal N}^i_{kl}$ are polynomials in $d$. Finally, 
 ${\cal M}^i_{kl}$ is the inverse of ${\cal N}^i_{kl}$ and then 
\begin{equation}
\Sigma^{{\cal O}^i}_{(k)}(p,q) ~=~ {\cal M}^i_{kl} 
{\cal P}^{i ~\, \mu_1 \ldots \mu_{n_i}}_{(l)}(p,q) \left. \left(  
\left\langle \psi(p) {\cal O}^i_{\mu_1 \ldots \mu_{n_i}}(-p-q) \bar{\psi}(q) 
\right\rangle \right) \right|_{p^2=q^2=-\mu^2} 
\end{equation}
where again there is no summation over the level label $i$.

\subsection{Scalar.}

\begin{equation}
{\cal P}^{S}_{(1)}(p,q) ~=~ \Gamma_{(0)} ~~~,~~~ 
{\cal P}^{S}_{(2)}(p,q) ~=~ \frac{1}{\mu^2} \Gamma_{(2)}^{pq} 
\end{equation}
\begin{equation}
{\cal M}^S ~=~ \frac{1}{12} \left(
\begin{array}{cc}
3 & 0 \\
0 & - 4 \\
\end{array}
\right) ~. 
\end{equation}
In order to compare with \cite{33}, the mapping between the amplitudes is 
\begin{equation}
\Sigma^{S}_{(1)}(p,q) ~=~ -~ A_S ~~~,~~~ \Sigma^{S}_{(2)}(p,q) ~=~ -~ 2 C_S 
\end{equation}
where the notation of the right hand side is that of \cite{33}.  

\subsection{Vector.}

\begin{eqnarray}
{\cal P}^{V}_{(1) \mu }(p,q) &=& \gamma_\mu ~~~,~~~ 
{\cal P}^{V}_{(2) \mu }(p,q) ~=~ \frac{{p}^\mu \pslash}{\mu^2} ~~~,~~~ 
{\cal P}^{V}_{(3) \mu }(p,q) ~=~ \frac{{p}_\mu \qslash}{\mu^2} ~, \nonumber \\
{\cal P}^{V}_{(4) \mu }(p,q) &=& \frac{{q}_\mu \pslash}{\mu^2} ~~~,~~~ 
{\cal P}^{V}_{(5) \mu }(p,q) ~=~ \frac{{q}_\mu \qslash}{\mu^2} ~~~,~~~ 
{\cal P}^{V}_{(6) \mu }(p,q) ~=~ \frac{1}{\mu^2} \Gamma_{(3) \, \mu p q} ~. 
\end{eqnarray}
\begin{equation}
{\cal M}^V ~=~ \frac{1}{36(d-2)} \left(
\begin{array}{cccccc}
9 & 12 & 6 & 6 & 12 & 0 \\
12 & 16 (d - 1) &  8 (d - 1) &  8 (d - 1) & 4 (d + 2) & 0 \\
6 & 8 (d - 1) & 4 (4 d - 7) &  4 (d - 1) & 8 (d - 1) & 0 \\
6 & 8 (d - 1) &  4 (d - 1) & 4 (4 d - 7) & 8 (d - 1) & 0 \\
12 & 4 (d + 2) &  8 (d - 1) &  8 (d - 1) & 16 (d - 1) & 0 \\
0 & 0 & 0 & 0 & 0 & - 12 \\
\end{array}
\right) ~. 
\end{equation}
For comparison with \cite{33} the amplitudes are related by 
\begin{eqnarray}
\Sigma^{V}_{(1)}(p,q) &=& -~ A_V ~+~ 2 B_V ~+~ \frac{1}{2} C_V ~+~ 
\frac{1}{2} D_V ~~,~~ 
\Sigma^{V}_{(2)}(p,q) ~=~ \Sigma^{V}_{(5)}(p,q) ~=~ 2 B_V ~, \nonumber \\
\Sigma^{V}_{(3)}(p,q) &=& \Sigma^{V}_{(4)}(p,q) ~=~ -~ C_V ~-~ D_V ~~,~~
\Sigma^{V}_{(6)}(p,q) ~=~ C_V ~-~ D_V ~.
\end{eqnarray}

\subsection{Tensor.}

\begin{eqnarray}
{\cal P}^{T}_{(1) \mu \nu }(p,q) &=& \Gamma_{(2) \, \mu\nu} ~~~,~~~ 
{\cal P}^{T}_{(2) \mu \nu }(p,q) ~=~ \frac{1}{\mu^2} \left[ p_\mu q_\nu - p_\nu
q_\mu \right] \Gamma_{(0)} ~, \nonumber \\  
{\cal P}^{T}_{(3) \mu \nu }(p,q) &=& \frac{1}{\mu^2} 
\left[ \Gamma_{(2) \, \mu p} p_\nu - \Gamma_{(2) \, \nu p} p_\mu 
\right] ~~~,~~~ 
{\cal P}^{T}_{(4) \mu \nu }(p,q) ~=~ \frac{1}{\mu^2} 
\left[ \Gamma_{(2) \, \mu p} q_\nu - \Gamma_{(2) \, \nu p} q_\mu 
\right] ~, \nonumber \\ 
{\cal P}^{T}_{(5) \mu \nu }(p,q) &=& \frac{1}{\mu^2} 
\left[ \Gamma_{(2) \, \mu q} p_\nu - \Gamma_{(2) \, \nu q} p_\mu 
\right] ~~~,~~~ 
{\cal P}^{T}_{(6) \mu \nu }(p,q) ~=~ \frac{1}{\mu^2} 
\left[ \Gamma_{(2) \, \mu q} q_\nu - \Gamma_{(2) \, \nu q} q_\mu 
\right] ~, \nonumber \\ 
{\cal P}^{T}_{(7) \mu \nu }(p,q) &=& \frac{1}{\mu^4} 
\left[ \Gamma_{(2) \, p q} p_\mu q_\nu - \Gamma_{(2) \, p q} p_\nu q_\mu 
\right] ~~~,~~~ 
{\cal P}^{T}_{(8) \mu \nu }(p,q) ~=~ \frac{1}{\mu^2} 
\Gamma_{(4) \, \mu \nu p q} ~. 
\end{eqnarray}
To save space for the projection matrix in this case, we have partitioned the
$8$~$\times$~$8$ matrix into four submatrices. We have  
\begin{equation}
{\cal M}^T ~=~ \frac{1}{36(d-2)(d-3)} \left(
\begin{array}{cc}
{\cal M}^T_{11} & {\cal M}^T_{12} \\ 
{\cal M}^T_{21} & {\cal M}^T_{22} \\ 
\end{array}
\right) ~, 
\end{equation}
\begin{eqnarray}
{\cal M}^T_{11} &=& \left(
\begin{array}{cccc}
- 9 & 0 & - 12 & - 6 \\
0 & 6 (d-2) (d-3) & 0 & 0 \\
- 12 & 0 & - 8 (d-1) & - 4 (d-1) \\
- 6 & 0 & - 4 (d-1) & - 4 (2d-5) \\ 
\end{array}
\right) ~, \nonumber \\ 
{\cal M}^T_{12} &=& \left(
\begin{array}{cccc}
- 6 & - 12 & - 12 & - 6 \\
0 & 0 & 0 & 0 \\
- 4 (d-1) & - 2 (d+5) & - 8 (d-1) & 0 \\
- 2 (d-1) & - 4 (d-1) & - 4 (d-1) & 0 \\ 
\end{array}
\right) ~, \nonumber \\ 
{\cal M}^T_{21} &=& \left(
\begin{array}{cccc}
- 6 & 0 & - 4 (d-1) & - 2 (d-1) \\
- 12 & 0 & - 2 (d+5) & - 4 (d-1) \\
- 12 & 0 & - 8 (d-1) & - 4 (d-1) \\
0 & 0 & 0 & 0 \\ 
\end{array}
\right) ~, \nonumber \\ 
{\cal M}^T_{22} &=& \left(
\begin{array}{cccc}
- 4 (2d-5) & - 4 (d-1) & - 4 (d-1) & 0 \\
- 4 (d-1) & - 8 (d-1)  & - 8 (d-1) & 0 \\
- 4 (d-1) & - 8 (d-1) & - 8 (d-1) (d-2) & 0 \\
0 & 0 & 0 & 12 \\ 
\end{array}
\right) ~. 
\end{eqnarray}
In the tensor case the relations with the amplitudes of \cite{33} are
\begin{eqnarray}
\Sigma^{T}_{(1)}(p,q) &=& -~ A_T ~+~ \frac{1}{2} C_T ~~,~~ 
\Sigma^{T}_{(2)}(p,q) ~=~ -~ 2 B_T ~+~ C_T ~, \nonumber \\
\Sigma^{T}_{(3)}(p,q) &=& \Sigma^{T}_{(6)}(p,q) ~=~ 2 C_T ~~,~~ 
\Sigma^{T}_{(4)}(p,q) ~=~ \Sigma^{T}_{(5)}(p,q) ~=~ -~ 2 B_T ~+~ C_T ~,
\nonumber \\  
\Sigma^{T}_{(8)}(p,q) &=& 2 B_T ~-~ C_T ~.
\end{eqnarray}

\subsection{Wilson $2$.}

\begin{eqnarray}
{\cal P}^{W_2}_{(1) \mu \nu }(p,q) &=& \gamma_\mu p_\nu + \gamma_\nu p_\mu
- \frac{2}{d} \pslash \eta_{\mu\nu} ~~,~~ 
{\cal P}^{W_2}_{(2) \mu \nu }(p,q) ~=~ \gamma_\mu q_\nu + \gamma_\nu q_\mu
- \frac{2}{d} \qslash \eta_{\mu\nu} ~, \nonumber \\
{\cal P}^{W_2}_{(3) \mu \nu }(p,q) &=& \pslash \left[ 
\frac{1}{\mu^2} p_\mu p_\nu + \frac{1}{d} \eta_{\mu\nu} \right] ~~,~~ 
{\cal P}^{W_2}_{(4) \mu \nu }(p,q) ~=~ \pslash \left[ 
\frac{1}{\mu^2} p_\mu q_\nu + \frac{1}{\mu^2} q_\mu p_\nu
- \frac{1}{d} \eta_{\mu\nu} \right] ~, \nonumber \\ 
{\cal P}^{W_2}_{(5) \mu \nu }(p,q) &=& \pslash \left[ 
\frac{1}{\mu^2} q_\mu q_\nu + \frac{1}{d} \eta_{\mu\nu} \right] ~~,~~ 
{\cal P}^{W_2}_{(6) \mu \nu }(p,q) ~=~ \qslash \left[ 
\frac{1}{\mu^2} p_\mu p_\nu + \frac{1}{d} \eta_{\mu\nu} \right] ~, 
\nonumber \\
{\cal P}^{W_2}_{(7) \mu \nu }(p,q) &=& \qslash \left[ 
\frac{1}{\mu^2} p_\mu q_\nu + \frac{1}{\mu^2} q_\mu p_\nu 
-\frac{1}{d} \eta_{\mu\nu} \right] ~,~
{\cal P}^{W_2}_{(8) \mu \nu }(p,q) ~=~ \qslash \left[ 
\frac{1}{\mu^2} q_\mu q_\nu + \frac{1}{d} \eta_{\mu\nu} \right] ~, \nonumber \\
{\cal P}^{W_2}_{(9) \mu \nu }(p,q) &=& \frac{1}{\mu^2} \left[ 
\Gamma_{(3) \, \mu p q } p_\nu + \Gamma_{(3) \, \nu p q } p_\mu \right] ~, 
\nonumber \\
{\cal P}^{W_2}_{(10) \mu \nu }(p,q) &=& \frac{1}{\mu^2} \left[ 
\Gamma_{(3) \, \mu p q } q_\nu + \Gamma_{(3) \, \nu p q } q_\mu \right] ~.
\end{eqnarray}
Similar to the previous case we have subdivided the $10$~$\times$~$10$ matrix
here into four submatrices in order to ease the presentation. We have 
\begin{eqnarray}
{\cal M}^{W_2} &=& -~ \frac{1}{108(d-2)} \left(
\begin{array}{cc}
{\cal M}^{W_2}_{11} & {\cal M}^{W_2}_{12} \\ 
{\cal M}^{W_2}_{21} & {\cal M}^{W_2}_{22} \\ 
\end{array}
\right) ~, \nonumber \\ 
{\cal M}^{W_2}_{11} &=& \left(
\begin{array}{ccccc}
18 & 9 & 48 & 24 & 12 \\ 
9 & 18 & 24 & 30 & 24 \\ 
48 & 24 & 64 (d+1) & 32 (d+1) & 16 (d+4) \\ 
24 & 30 & 32 (d+1) & 8 (5d-1) & 8 (4d+1) \\ 
12 & 24 & 16 (d+4) & 8 (4d+1) & 32 (2d-1) \\
\end{array}
\right) ~, \nonumber \\ 
{\cal M}^{W_2}_{12} &=& \left(
\begin{array}{ccccc}
24 & 30 & 24 & 0 & 0 \\ 
12 & 24 & 48 & 0 & 0 \\ 
32 (d+1) & 16 (d+4) & 8 (d+10) & 0 & 0 \\ 
16 (d+1) & 20 (d+1) & 16 (d+4) & 0 & 0 \\ 
8 (d+4) & 16 (d+1) & 32 (d+1) & 0 & 0 \\ 
\end{array}
\right) ~, \nonumber \\ 
{\cal M}^{W_2}_{21} &=& \left(
\begin{array}{ccccc}
24 & 12 & 32 (d+1) & 16 (d+1) & 8 (d+4) \\ 
30 & 24 & 16 (d+4) & 20 (d+1) & 16 (d+1) \\ 
24 & 48 & 8 (d+10) & 16 (d+4) & 32 (d+1) \\ 
0 & 0 & 0 & 0 & 0 \\
0 & 0 & 0 & 0 & 0 \\
\end{array}
\right) ~, \nonumber \\ 
{\cal M}^{W_2}_{22} &=& \left(
\begin{array}{ccccc}
32 (2d-1) & 8 (4d+1) & 16 (d+4) & 0 & 0 \\ 
8 (4d+1) & 8 (5d-1) & 32 (d+1) & 0 & 0 \\ 
16 (d+4) & 32 (d+1) & 64 (d+1) & 0 & 0 \\ 
0 & 0 & 0 & - 24 & - 12 \\
0 & 0 & 0 & - 12 & - 24 \\
\end{array}
\right) ~. 
\end{eqnarray}

\subsection{Wilson $3$.}

\begin{eqnarray}
{\cal P}^{W_3}_{(1) \mu \nu \sigma}(p,q) &=& \frac{1}{\mu^2} \left[
\gamma_\mu p_\nu p_\sigma + \gamma_\nu p_\sigma p_\mu
+ \gamma_\sigma p_\mu p_\nu \right] ~+~ \frac{1}{[d+2]} \left[ 
\gamma_\mu \eta_{\nu\sigma} + \gamma_\nu \eta_{\sigma\mu}
+ \gamma_\sigma \eta_{\mu\nu} \right] \nonumber \\
&& -~ \frac{2\pslash}{[d+2]\mu^2} \left[ \eta_{\mu\nu} p_\sigma
+ \eta_{\nu\sigma} p_\mu + \eta_{\sigma\mu} p_\nu \right] ~, \nonumber \\ 
{\cal P}^{W_3}_{(2) \mu \nu \sigma}(p,q) &=& \frac{1}{\mu^2} \left[
\gamma_\mu p_\nu q_\sigma + \gamma_\nu p_\sigma q_\mu
+ \gamma_\sigma p_\mu q_\nu + \gamma_\mu q_\nu p_\sigma 
+ \gamma_\nu q_\sigma p_\mu + \gamma_\sigma q_\mu p_\nu \right] \nonumber \\
&& -~ \frac{1}{[d+2]} \left[ 
\gamma_\mu \eta_{\nu\sigma} + \gamma_\nu \eta_{\sigma\mu}
+ \gamma_\sigma \eta_{\mu\nu} \right] ~-~ \frac{2\pslash}{[d+2]\mu^2} \left[ 
\eta_{\mu\nu} q_\sigma + \eta_{\nu\sigma} q_\mu + \eta_{\sigma\mu} q_\nu 
\right] \nonumber \\ 
&& -~ \frac{2\qslash}{[d+2]\mu^2} \left[ \eta_{\mu\nu} p_\sigma
+ \eta_{\nu\sigma} p_\mu + \eta_{\sigma\mu} p_\nu \right] ~, \nonumber \\ 
{\cal P}^{W_3}_{(3) \mu \nu \sigma}(p,q) &=& \frac{1}{\mu^2} \left[
\gamma_\mu q_\nu q_\sigma + \gamma_\nu q_\sigma q_\mu
+ \gamma_\sigma q_\mu q_\nu \right] ~+~ \frac{1}{[d+2]} \left[ 
\gamma_\mu \eta_{\nu\sigma} + \gamma_\nu \eta_{\sigma\mu}
+ \gamma_\sigma \eta_{\mu\nu} \right] \nonumber \\
&& -~ \frac{2\qslash}{[d+2]\mu^2} \left[ \eta_{\mu\nu} q_\sigma
+ \eta_{\nu\sigma} q_\mu + \eta_{\sigma\mu} q_\nu \right] ~, \nonumber \\ 
{\cal P}^{W_3}_{(4) \mu \nu \sigma}(p,q) &=& \frac{\pslash}{\mu^4} 
p_\mu p_\nu p_\sigma ~+~ \frac{\pslash}{[d+2]\mu^2} \left[ 
\eta_{\mu\nu} p_\sigma + \eta_{\nu\sigma} p_\mu + \eta_{\sigma\mu} p_\nu 
\right] ~, \nonumber \\
{\cal P}^{W_3}_{(5) \mu \nu \sigma}(p,q) &=& \frac{\pslash}{\mu^4} \left[ 
p_\mu p_\nu q_\sigma + p_\mu q_\nu p_\sigma + q_\mu p_\nu p_\sigma \right] 
\nonumber \\
&& -~ \frac{\pslash}{[d+2]\mu^2} \left[ \eta_{\mu\nu} p_\sigma 
- \eta_{\mu\nu} q_\sigma + \eta_{\nu\sigma} p_\mu - \eta_{\nu\sigma} q_\mu 
+ \eta_{\sigma\mu} p_\nu - \eta_{\sigma\mu} q_\nu \right] ~, \nonumber \\ 
{\cal P}^{W_3}_{(6) \mu \nu \sigma}(p,q) &=& \frac{\pslash}{\mu^4} \left[ 
p_\mu q_\nu q_\sigma + q_\mu p_\nu q_\sigma + q_\mu q_\nu p_\sigma \right] 
\nonumber \\
&& +~ \frac{\pslash}{[d+2]\mu^2} \left[ \eta_{\mu\nu} p_\sigma 
- \eta_{\mu\nu} q_\sigma + \eta_{\nu\sigma} p_\mu - \eta_{\nu\sigma} q_\mu 
+ \eta_{\sigma\mu} p_\nu - \eta_{\sigma\mu} q_\nu \right] ~, \nonumber \\ 
{\cal P}^{W_3}_{(7) \mu \nu \sigma}(p,q) &=& \frac{\pslash}{\mu^4} 
q_\mu q_\nu q_\sigma ~+~ \frac{\pslash}{[d+2]\mu^2} \left[ 
\eta_{\mu\nu} q_\sigma + \eta_{\nu\sigma} q_\mu + \eta_{\sigma\mu} q_\nu 
\right] ~, \nonumber \\
{\cal P}^{W_3}_{(8) \mu \nu \sigma}(p,q) &=& \frac{\qslash}{\mu^4} 
p_\mu p_\nu p_\sigma ~+~ \frac{\qslash}{[d+2]\mu^2} \left[ 
\eta_{\mu\nu} p_\sigma + \eta_{\nu\sigma} p_\mu + \eta_{\sigma\mu} p_\nu 
\right] ~, \nonumber \\
{\cal P}^{W_3}_{(9) \mu \nu \sigma}(p,q) &=& \frac{\qslash}{\mu^4} \left[ 
p_\mu p_\nu q_\sigma + p_\mu q_\nu p_\sigma + q_\mu p_\nu p_\sigma \right] 
\nonumber \\
&& -~ \frac{\qslash}{[d+2]\mu^2} \left[ \eta_{\mu\nu} p_\sigma 
- \eta_{\mu\nu} q_\sigma + \eta_{\nu\sigma} p_\mu - \eta_{\nu\sigma} q_\mu 
+ \eta_{\sigma\mu} p_\nu - \eta_{\sigma\mu} q_\nu \right] ~, \nonumber \\ 
{\cal P}^{W_3}_{(10) \mu \nu \sigma}(p,q) &=& \frac{\qslash}{\mu^4} \left[ 
p_\mu q_\nu q_\sigma + q_\mu p_\nu q_\sigma + q_\mu q_\nu p_\sigma \right] 
\nonumber \\
&& +~ \frac{\qslash}{[d+2]\mu^2} \left[ \eta_{\mu\nu} p_\sigma 
- \eta_{\mu\nu} q_\sigma + \eta_{\nu\sigma} p_\mu - \eta_{\nu\sigma} q_\mu 
+ \eta_{\sigma\mu} p_\nu - \eta_{\sigma\mu} q_\nu \right] ~, \nonumber \\ 
{\cal P}^{W_3}_{(11) \mu \nu \sigma}(p,q) &=& \frac{\qslash}{\mu^4} 
q_\mu q_\nu q_\sigma ~+~ \frac{\qslash}{[d+2]\mu^2} \left[ 
\eta_{\mu\nu} q_\sigma + \eta_{\nu\sigma} q_\mu + \eta_{\sigma\mu} q_\nu 
\right] ~, \nonumber \\
{\cal P}^{W_3}_{(12) \mu \nu \sigma}(p,q) &=& \frac{1}{\mu^4} \left[ 
\Gamma_{(3) \, \mu p q} p_\nu p_\sigma +
\Gamma_{(3) \, \nu p q} p_\mu p_\sigma +
\Gamma_{(3) \, \sigma p q} p_\mu p_\nu \right] \nonumber \\
&& +~ \frac{1}{[d+2]\mu^2} \left[ 
\Gamma_{(3) \, \mu p q} \eta_{\nu\sigma} +
\Gamma_{(3) \, \nu p q} \eta_{\mu\sigma} +
\Gamma_{(3) \, \sigma p q} \eta_{\mu\nu} \right] ~, \nonumber \\ 
{\cal P}^{W_3}_{(13) \mu \nu \sigma}(p,q) &=& \frac{1}{\mu^4} \left[ 
\Gamma_{(3) \, \mu p q} p_\nu q_\sigma +
\Gamma_{(3) \, \nu p q} p_\mu q_\sigma +
\Gamma_{(3) \, \sigma p q} p_\mu q_\nu \right. \nonumber \\
&& \left. ~~~~+~ \Gamma_{(3) \, \mu p q} q_\nu p_\sigma +
\Gamma_{(3) \, \nu p q} q_\mu p_\sigma +
\Gamma_{(3) \, \sigma p q} q_\mu p_\nu \right] \nonumber \\
&& -~ \frac{1}{[d+2]\mu^2} \left[ 
\Gamma_{(3) \, \mu p q} \eta_{\nu\sigma} +
\Gamma_{(3) \, \nu p q} \eta_{\mu\sigma} +
\Gamma_{(3) \, \sigma p q} \eta_{\mu\nu} \right] ~, \nonumber \\ 
{\cal P}^{W_3}_{(14) \mu \nu \sigma}(p,q) &=& \frac{1}{\mu^4} \left[ 
\Gamma_{(3) \, \mu p q} q_\nu q_\sigma +
\Gamma_{(3) \, \nu p q} q_\mu q_\sigma +
\Gamma_{(3) \, \sigma p q} q_\mu q_\nu \right] \nonumber \\
&& +~ \frac{1}{[d+2]\mu^2} \left[ 
\Gamma_{(3) \, \mu p q} \eta_{\nu\sigma} +
\Gamma_{(3) \, \nu p q} \eta_{\mu\sigma} +
\Gamma_{(3) \, \sigma p q} \eta_{\mu\nu} \right] ~. 
\end{eqnarray}
Here the partition of the $14$~$\times$~$14$ matrix is into ten non-zero
submatrices. We have 
\begin{eqnarray}
{\cal M}^{W_3} &=& \frac{1}{2106d(d-2)} \left(
\begin{array}{cccc}
{\cal M}^{W_3}_{11} & {\cal M}^{W_3}_{12} & {\cal M}^{W_3}_{13} & 0 \\
{\cal M}^{W_3}_{21} & {\cal M}^{W_3}_{22} & {\cal M}^{W_3}_{23} & 0 \\
{\cal M}^{W_3}_{31} & {\cal M}^{W_3}_{32} & {\cal M}^{W_3}_{33} & 0 \\
0 & 0 & 0 & {\cal M}^{W_3}_{44} \\
\end{array}
\right) ~, \nonumber 
\end{eqnarray}
where the $\Gamma_{(3)}$ sector corresponds to the lower outer corner submatrix
or subspace.
\begin{eqnarray}
{\cal M}^{W_3}_{11} &=& \left(
\begin{array}{cccc}
312 (d+1) & 156 (d+1) & 78 (d+4) & 1248 (d+1) \\ 
156 (d+1) & 39 (5d+2) & 156 (d+1) & 624 (d+1) \\ 
78 (d+4) & 156 (d+1) & 312 (d+1) & 312 (d+4) \\ 
1248 (d+1) & 624 (d+1) & 312 (d+4) & 1664 (d+3) (d+1) \\ 
\end{array}
\right) ~, \nonumber \\ 
{\cal M}^{W_3}_{12} &=& \left(
\begin{array}{cccc}
\! 624 (d+1) & \! 312 (d+2) & 156 (d+4) & 624 (d+1) \\ 
\! 312 (2d+1) & \! 156 (3d+2) & 312 (d+1) & 312 (d+1) \\ 
\! 156 (3d+4) & \! 624 (d+1) & 624 (d+1) & 156 (d+4) \\ 
\! 832 (d+3) (d+1) & \! 416 (d+6) (d+1) & 208 (d+12) (d+1) & 832 (d+3) (d+1) \\ 
\end{array}
\right) ~, \nonumber \\ 
{\cal M}^{W_3}_{13} &=& \left(
\begin{array}{ccc}
624 (d+1) & 156 (3d+4) & 312 (d+4) \\ 
156 (3d+2) & 312 (2d+1) & 624 (d+1) \\ 
312 (d+2) & 624 (d+1) & 1248 (d+1) \\ 
416 (d+6) (d+1) & 208 (d+12) (d+1) & 104 (d^2+22d+48) \\ 
\end{array}
\right) ~, \nonumber \\ 
{\cal M}^{W_3}_{21} &=& \left(
\begin{array}{cccc}
624 (d+1) & 312 (2d+1) & 156 (3d+4) & 832 (d+3) (d+1) \\ 
312 (d+2) & 156 (3d+2) & 624 (d+1) & 416 (d+6) (d+1) \\ 
156 (d+4) & 312 (d+1) & 624 (d+1) & 208 (d+12) (d+1) \\ 
624 (d+1) & 312 (d+1) & 156 (d+4) & 832 (d+3) (d+1) \\ 
\end{array}
\right) ~, \nonumber \\ 
{\cal M}^{W_3}_{22} &=& (d+1) \left(
\begin{array}{cccc}
416 (2d+3) & 624 (d+2) & 104 \frac{(4d^2+25d+12)}{(d+1)} & 416 (d+3) \\ 
624 (d+2) & 208 \frac{(4d^2+7d+6)}{(d+1)} & 416 (2d+3) & 208 (d+6) \\ 
104 \frac{(4d^2+25d+12)}{(d+1)} & 416 (2d+3) & 416 (4d+3) & 104 (d+12) \\ 
416 (d+3) & 208 (d+6) & 104 (d+12) & 416 (4d+3) \\ 
\end{array}
\right) ~, \nonumber \\ 
{\cal M}^{W_3}_{23} &=& \left(
\begin{array}{ccc}
416 (d+3) (d+1) & 312 (d^2+7d+4) & 208 (d+12) (d+1) \\ 
312 (d+2)^2 & 416 (d+3) (d+1) & 416 (d+6) (d+1) \\ 
208 (d+6) (d+1) & 416 (d+3) (d+1) & 832 (d+3) (d+1) \\ 
416 (2d+3) (d+1) & 104 (4d^2+25d+12) & 208 (d+12) (d+1) \\ 
\end{array}
\right) ~, \nonumber \\ 
{\cal M}^{W_3}_{31} &=& \left(
\begin{array}{cccc}
624 (d+1) & 156 (3d+2) & 312 (d+2) & 416 (d+6) (d+1) \\ 
156 (3d+4) & 312 (2d+1) & 624 (d+1) & 208 (d+12) (d+1) \\ 
312 (d+4) & 624 (d+1) & 1248 (d+1) & 104 (d^2+22d+48) \\ 
\end{array}
\right) ~, \nonumber \\ 
{\cal M}^{W_3}_{32} &=& \left(
\begin{array}{cccc}
\! 416 (d+3) (d+1) \! & \! 312 (d+2)^2 \! & \! 208 (d+6) (d+1) & 
\! \! 416 (2d+3) (d+1) \! \\ 
\! 312 (d^2+7d+4) \! & \! 416 (d+3) (d+1) \! & \! 416 (d+3) (d+1) & 
\! \! 104 (4d^2+25d+12) \! \\ 
\! 208 (d+12) (d+1) \! & \! 416 (d+6) (d+1) \! & \! 832 (d+3) (d+1) & 
\! \! 208 (d+12) (d+1) \! \\
\end{array}
\right) \,, \nonumber \\ 
{\cal M}^{W_3}_{33} &=& \left(
\begin{array}{ccc}
208 (4d^2+7d+6) & 624 (d+2) (d+1) & 416 (d+6) (d+1) \\ 
624 (d+2) (d+1) & 416 (2d+3) (d+1) & 832 (d+3) (d+1) \\ 
416 (d+6) (d+1) & 832 (d+3) (d+1) & 1664 (d+3) (d+1) \\ 
\end{array}
\right) \,, \nonumber \\
{\cal M}^{W_3}_{44} &=& \left(
\begin{array}{ccc}
- 416 (d+1) & - 208 (d+1) & - 104 (d+4) \\
- 208 (d+1) & - 52 (5d+2) & - 208 (d+1) \\ 
- 104 (d+4) & - 208 (d+1) & - 416 (d+1) \\ 
\end{array}
\right) \,. 
\end{eqnarray}

\end{document}